\documentclass[10pt]{iopart}
\pdfoutput=1
\usepackage{graphicx}

\eqnobysec
\newcommand{\beq}{\begin{equation}}
\newcommand{\beqa}{\begin{eqnarray}}
\newcommand{\eeq}{\end{equation}}
\newcommand{\eeqa}{\end{eqnarray}}

\renewcommand{\d}{{\rm d }}

\renewcommand{\i}{{\rm i}}
\newcommand{\mean}[1]{\langle#1\rangle}

\renewcommand{\c}{{\cal C}}

\newcommand{\s}{{\sigma}}

\newcommand{\w}{{\bar w}}
\newcommand{\F}{\mathrm{ F}}
\renewcommand{\L}{\mathrm{ L}_\tau}
\newcommand{\Ls}{\mathrm{ L}_s}
\newcommand{\ps}{u}
\newcommand{\eq}{\mathrm{eq}}
\newcommand{\stat}{\mathrm{stat}}

\begin{document}

\title{Dynamics of the directed Ising chain}
\date{\today}
\author{Claude Godr\`eche}
\address{
Institut de Physique Th\'eorique, CEA Saclay and CNRS,\\
91191 Gif-sur-Yvette cedex, France}
\begin{abstract}
The study by Glauber of the time-dependent statistics of the Ising chain is extended to the case where each spin is influenced unequally by its nearest neighbours.
The asymmetry of the dynamics implies the failure of the detailed balance condition.
The functional form of the rate at which an individual spin changes its state is constrained by the global balance condition
with respect to the equilibrium measure of the Ising chain.
The local magnetization, the equal-time and two-time correlation functions and the linear response to an external magnetic field obey linear equations which are solved explicitly.
The behaviour of these quantities and the relation between the correlation and response functions are analyzed both in the stationary state and in the zero-temperature scaling regime.
In the stationary state, a transition between two behaviours of the correlation function occurs when the amplitude of the asymmetry crosses a critical value, with the consequence that the
limit fluctuation-dissipation ratio decays continuously from the value 1, for the equilibrium state in the absence of asymmetry, to 0 for this critical value.
At zero temperature, under asymmetric dynamics, the system loses its critical character, yet keeping many of the characteristic features of a coarsening system.
\end{abstract}

\maketitle

\section{Introduction}

In his classical work on the time-dependent statistics of the Ising model~\cite{glau} Glauber addresses the question of how to define the dynamics of a one-dimensional chain of $N$ spins, 
$\s_n(t)=\pm1$, 
evolving from an arbitrary initial condition, in such a way that equilibrium is approached at large times. 
The spins, in contact with a heat reservoir, make transitions randomly between the two values $\pm1$.
The rate, or probability per unit time, at which an individual spin flips is also assumed to depend on the values of its neighbouring spins.
The choice made in~\cite{glau} for the rate at which the $n$-th spin flips from the value $\s_n$ to $-\s_n$, the other remaining fixed, is
\beq\label{eq:glau}
w(\s_n)=\frac{1}{2}\alpha\big[1-\frac{1}{2}\gamma\s_n(\s_{n-1}+\s_{n+1})\big].
\eeq
The parameter $\alpha$ appearing in this expression gives a time scale, while
$\gamma$ is, for the time being, a free parameter.

There are three possible values taken by the rate function~(\ref{eq:glau}) associated to
the eight configurations of the group of spins $\{\s_n;\s_{n-1},\s_{n+1}\}$ involved in the flipping of the spin $\s_n$,
\beq\label{three}
\frac{1}{2}\alpha(1-\gamma),\qquad\frac{1}{2}\alpha,\qquad
\frac{1}{2}\alpha(1+\gamma).
\eeq
Assuming that $\gamma$ is positive, the configurations $\{+;++\}$ and $\{-;--\}$, where $\s_n$ is parallel to its neighbours, are longer-lived than the configurations $\{+;--\}$ and $\{-;++\}$ where it is antiparallel since these configurations correspond respectively to the rates 
$\frac{1}{2}\alpha(1-\gamma)$ and $\frac{1}{2}\alpha(1+\gamma)$.
The intermediate rate $\frac{1}{2}\alpha$ corresponds to the cases where the neighbouring spins are antiparallel, $\{+;+-\}$, $\{+;-+\}$, $\{-;+-\}$, $\{-;-+\}$.
The parameter $\gamma$ thus describes the tendency of spins toward ferromagnetic alignment and determines the equilibrium state towards which the model defined by the expression~(\ref{eq:glau}) of the rate relaxes.
The choice $\gamma<0$ would describe the antiferromagnetic case.

The correspondence of the stochastic model defined so far with the Ising chain with energy function (or Hamiltonian)
\beq\label{hamilt}
E({\cal C})=-J\sum_n\s_n\s_{n+1},
\eeq
$J$ denoting the coupling constant ($J>0$), and ${\c}=\{\s_1,\ldots,\s_N\}$ a spin configuration,
is made by requesting that the rate~(\ref{eq:glau}) satisfy the 
detailed balance condition at equilibrium~\cite{glau},
\beq
P(\c)w(\s_n)=P(\c_{n})w(-\s_n),
\eeq
where the weight $P(\c)$ is proportional to the Boltzmann factor $\exp(-E(\c)/T)$,
and $\c_n$ denotes the configuration obtained from $\c$ by flipping $\s_n$.
Denoting by $K=J/T$ the reduced coupling constant, this condition leads to the single constraint relation on $\gamma$,
\beq\label{eq:db}
\frac{1}{2}\alpha(1+\gamma)=\e^{4K}\frac{1}{2}\alpha(1-\gamma),
\eeq
which yields~\cite{glau}
\beq\label{gam}
\gamma=\tanh 2K.
\eeq

With this value of the parameter $\gamma$ the rate~(\ref{eq:glau}) now ensures that the system prepared in an arbitrary initial condition relaxes to the equilibrium state of the Ising model~(\ref{hamilt}).
The sequel of ref.~\cite{glau} is devoted to the complete determination of the time-dependence of the magnetization, of the equal-time two-spin correlation function, of the two-time two-spin correlation function, and finally of the quantities relevant for a description of the model in the presence of a uniform time-varying magnetic field.
The solvability of the model relies on the linearity of the equations giving the temporal evolution of the quantities of interest, itself due to the choice of the rate~(\ref{eq:glau}).

The kinetic Ising model introduced by Glauber, defined by dynamical rules obeying detailed balance and relaxing from an arbitrary initial condition towards equilibrium, is probably the first and one of the simplest cases of a solvable model of nonequilibrium statistical mechanics of a strongly interacting system.
This model has therefore, ever since, been considered as a paradigm for the investigation of new aspects of nonequilibrium statistical mechanics. 
Recent significant advances in this field concern phase ordering at low-temperature~\cite{langer,bray},
where the one-dimensional kinetic Ising model with symmetric dynamics at zero temperature (or more generally in the low temperature scaling regime) appears as
the simplest example of a coarsening spin system~\cite{bray}.
Indeed, starting from a random initial condition (e.g. if the system was initially at equilibrium at infinite temperature),
the system is unable to relax to any of its two
ferromagnetically ordered, symmetry-related, equilibrium states, because the time to reach equilibrium can be arbitrarily large. 
On the contrary, domains of positive and negative magnetization grow forever,
the system stays perpetually out of equilibrium,
and, in the scaling regime, the system becomes statistically self-similar
with only one characteristic length scale, the mean size of domains.
Furthermore the two-time correlation and response functions 
only depend on the ratio of their arguments. 
The corresponding scaling functions are now well known, both for correlations~\cite{bray,cox,bray89,amar90,bray97,prados97,gl2000}, and for the response~\cite{gl2000,lip2000, malte,corb1}.
Extensions to multispin two-time correlation and response functions can be found in~\cite{sollich1}.

In the present work we address the situation of a {\it directed} Ising chain where the flipping spin is not equally influenced by its neighbours.
As for the Glauber-Ising model the dynamics is non conserved. 
The asymmetry of the dynamics means that
the detailed balance condition is not obeyed, and therefore there is no conventional 
equilibrium --one speaks instead of a stationary state.
A notable remark is that this stationary state does not carry any conserved macroscopic current, in contrast with what is usually met for the driven systems of current interest such as the Asymmetric Simple Exclusion Process~\cite{asep}, the Katz, Lebowitz and Spohn model~\cite{kls}, or the Zero Range Process~\cite{zrp}\footnote{For example, in the one-dimensional KLS model~\cite{kls,lg2006}, defined as an Ising chain with conserved
dynamics and totally asymmetric rules, only the $+-$ bond 
is updated, not the $-+$ bond.
Therefore if the $+$ spin is considered as a particle and the $-$ spin as a hole, matter is transported in the positive direction.}.

The definition of the model encompasses the partially asymmetric case, the totally asymmetric one, and the Glauber symmetric case, through a single parameter, the bias.
For any value of the bias the stationary state of the model is the same, namely, the equilibrium state of the symmetric case~\cite{gb2009}.
However the relaxation properties of the model toward its stationary state, as well as the nature of the fluctuations in this stationary state, depend on the value of the bias.

Kinetic Ising models with asymmetric conserved dynamics, such as the KLS model~\cite{kls}, represent a long-time established field of study~\cite{kls+}.
In constrast with the former, very few works have been devoted to the case of 
Ising models with non conserved asymmetric dynamics.
These models have been considered in the past~\cite{kun} and more recently~\cite{stau,gb2009}\footnote{Let us also mention a very recent publication devoted to the study of the algebraic properties of a disordered totally asymmetric Glauber model~\cite{ayyer}.}.
Ref.~\cite{kun} gives an example of a rate function for the two-dimensional kinetic Ising model with totally asymmetric dynamics (up to a seemingly missing factor 2) leading to a Gibbsian stationary measure.
However this work does not provide a systematic approach for the derivation of such forms of the rate,
nor general prescription for writing its expression.
Ref.~\cite{stau} is concerned with the numerical investigation of the possible existence of a phase transition for the kinetic Ising model in dimensions two to five if the usual Glauber form of the rate is modified by truncating the local field, keeping the influential spins only.
This rate neither fulfills detailed nor global balance, and therefore leads to an unknown stationary measure.
These two references motivated the work presented in~\cite{gb2009}.
The main focus was on the question: what is the general prescription for finding a rate such that the stationary state be Gibbsian, especially for totally asymmetric dynamics? 
In what dimension, if any, does such a rate exist?
The other question posed, inspired by~\cite{stau}, was:
does the Ising model augmented by Glauber dynamics with the truncated rate defined above exhibit a phase transition to a ferromagnetic state?

At zero temperature an alternative description of the model is in terms of diffusing and annihilating domain walls submitted to a bias, as explained later.
This situation has been explored in the past~\cite{gunter,stinch}.
A discussion of the results obtained in these studies will be given in the course of the present work.

The aim of the present study is to investigate, on the particular model defined below, the changes induced by directedness on the dynamics of the one-dimensional Glauber-Ising model.
The agenda for the first part of the work parallels that followed in~\cite{glau}.
We begin by making a choice of rates for the transitions experienced by the flipping spin,
in order to define the rules of the dynamics.
As in the case of symmetric dynamics, we want to keep a correspondence between the stochastic model and the Ising chain with Hamiltonian~(\ref{hamilt}).
Though one can no longer rely on the detailed balance condition to fix the form of the rate function,
it is instead possible to rely on the global balance condition to this end.
As shown in~\cite{gb2009}, it is possible to find rates such that the weight of configurations at stationarity is still given by the usual Boltzmann-Gibbs prescription for the Hamiltonian~(\ref{hamilt}).
The model thus defined relaxes at long times towards a stationary state whose measure is identical to that of the equilibrium state of the symmetric Ising-Glauber model~\cite{gb2009}.
We shall first recall the results of ref.~\cite{gb2009} concerning the determination of the rates leading to a stationary Boltzmann-Gibbs measure corresponding to the Hamiltonian~(\ref{hamilt}).
We shall then extend the study of ref.~\cite{gb2009} to the case of a chain submitted to an external magnetic field.

Having defined the process, we shall want to solve for the time dependence of the magnetization, and of the correlation and response functions for the directed Ising chain.
Fortunately our choice of rate~(\ref{direct}) still leads to linear equations, which makes the model solvable.
The novelty comes from the introduction of another scale, through the occurrence of a bias (or velocity).
Moreover, the absence of detailed balance, due to the asymmetry of the dynamics, implies the violation of the fluctuation-dissipation relation, valid at equilibrium for the time-reversible kinetic Ising model with symmetric dynamics~\cite{glau}.
We shall investigate the fate of the relation between the correlation and response functions in such circumstances.

Finally, paralleling the study done for the case of symmetric dynamics~\cite{gl2000},
we shall emphasize the low-temperature scaling behaviour of the model.
For the directed Ising chain, coarsening is still present but it is biased.
We shall present an analysis of this coarsening process and of the scaling behaviour of the quantities of interest mentioned above.
While for the symmetric time-reversible dynamics, the fluctuation-dissipation relation is violated at low temperature due to the existence of an infinite relaxation time $\tau_{\eq}$, here this relation is already violated in the stationary state, as said above.
We shall investigate the fluctuation-dissipation relation (in particular the behaviour of the fluctuation-dissipation ratio) in the zero-temperature regime where $\tau_{\eq}$ is very large, and thus where the system is doubly a nonequilibrium system: its dynamics violates detailed balance and it is `out of stationarity'.

\section{Definition of the rates for the directed Ising chain}

We consider the situation where the flipping spin $\s_n$ is unequally influenced by its left and right neighbours. 
We shall define the rates defining the dynamical rules such that they lead to a stationary state measure identical to the equilibrium Boltzmann-Gibbs measure for the Hamiltonian~(\ref{hamilt}), even though the dynamics is not reversible and therefore the detailed balance condition is violated in this stationary state.
The general study leading to the determination of such rates is given in~\cite{gb2009}.
For the case of the chain with Hamiltonian~(\ref{hamilt}) we shall merely state and explain the results. 
The case of the chain submitted to a magnetic field, which was not considered in~\cite{gb2009}, will be explained in greater detail.

\subsection{Rates in the absence of a magnetic field}

In the present work we use the following simple form of the rate function for the dynamics of the directed Ising chain,
\beq\label{direct}
w(\s_n)=\frac{1}{2}\alpha \left[1-\gamma\s_n(p\s_{n-1}+(1-p)\s_{n+1})\right ],
\eeq
where $0\le p\le 1$ is a parameter which quantifies the respective influence of the neighbours on the flipping spin.
For instance, $p=0$ ($p=1$) means that $\s_n$ only feels its right (left) neighbour, and $p=1/2$ corresponds to the symmetric Glauber form~(\ref{eq:glau}). 

There are now four possible values taken by the rate function~(\ref{direct}) associated to
the eight configurations of the group of spins $\{\s_n;\s_{n-1},\s_{n+1}\}$ involved in the flipping of the spin $\s_n$,
\beq
\frac{1}{2}\alpha(1-\gamma),\quad\frac{1}{2}\alpha(1-\gamma V),
\quad\frac{1}{2}\alpha(1+\gamma V),\quad
\frac{1}{2}\alpha(1+\gamma),
\eeq
where we define a velocity associated to the bias $p$ and denoted by $V=2p-1$.
Compared to the set of rates for symmetric dynamics given in~(\ref{three}) the only changes occur for the two intermediate rates where the neighbours of the flipping spin are antiparallel.

\begin{figure}
\begin{center}
\includegraphics[angle=0,width=.9\linewidth]{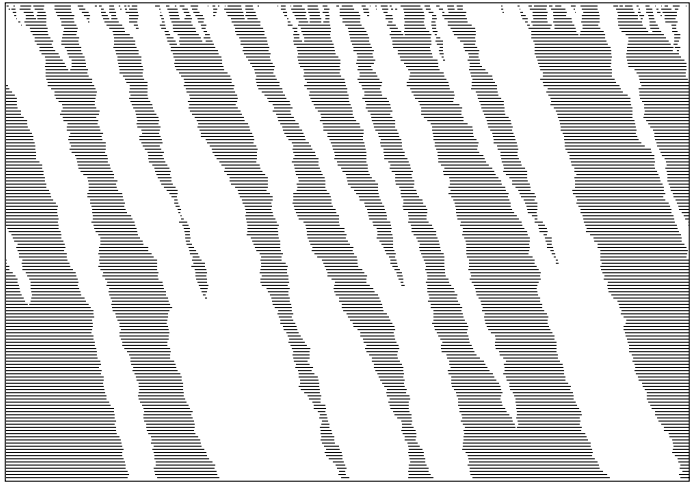}
\caption{\label{fig:visu}
Space-time representation of the zero-temperature biased coarsening of the directed Ising chain with periodic boundary conditions (drift velocity $V=0.4$). }
\end{center}
\end{figure}

The situation at zero temperature (or $\gamma=1$) is simpler because the rate $\frac{1}{2}\alpha(1-\gamma)$ corresponding to the occurrence of elementary thermal excitations ($+++\rightarrow +-+$) and ($---\rightarrow-+-$) vanishes. 
Let us fix $\alpha=1$, for simplicity.
The dynamics of the chain can entirely be described by the motion of domain walls. 
This is a well-known fact for the usual symmetric dynamics; each domain wall performs a symmetric random walk, and when two domain walls encounter they annihilate. 
In the present case of directed dynamics the domain walls perform a biased random walk, with rate $p$ for a step to the right and $1-p$ to the left, and they again annihilate when they come into contact (see Figure~\ref{fig:visu}). 
Indeed, steps to the right correspond to flipping the spin $\s_n$ in configurations $\{+;-+\}$ and $\{-;+-\}$,
with rate $\frac{1}{2}(1+ V)=p$.
Steps to the left correspond to flipping the spin $\s_n$ in configurations $\{+;+-\}$ and $\{-;-+\}$, with rate $\frac{1}{2}(1-V)=1-p$.
If $V>0$ the domain walls drift away to the right, because the latter configurations are longer lived than the former ones.
Hence, at zero temperature, the drift velocity of the domain walls is precisely the velocity $V$ defined above.
The annihilation of two domain walls corresponds to flipping the spin $\s_n$ in configurations $\{+;--\}$ and $\{-;++\}$, with rate $\frac{1}{2}\alpha(1+\gamma)=1$.

As shown below, the form of the rate~(\ref{direct}) fulfills our requirement of leading to a stationary state measure identical to the equilibrium Boltzmann-Gibbs measure for the Hamiltonian~(\ref{hamilt}).

%
%
\subsection{Constraint equations and derivation of (\ref{direct})}

In order to understand the origin of the expression~(\ref{direct}), we briefly come back on the main steps of the determination of the rates with the methods of~\cite{gb2009}.
The aim is to derive a set of constraints between the transition rates that need
to be satisfied in order for the stationary state measure to be Gibbsian. 

The dynamics consist in flipping a spin, chosen at random, say spin $n$, with a rate 
$w(\c_n|\c)$, which is a more formal notation for $w(\s_n)$, corresponding to the transition between configurations $\c=\{\s_1,\ldots,\s_{n},\ldots,\s_{N}\}$ and 
$\c_n=\{\s_1,\ldots,-\s_{n},\ldots,\s_{N}\}$.
We choose periodic boundary conditions.
At stationarity, the master equation expresses that losses are equal to gains, and reads
\beq\label{master0}
P(\c)\sum_{n}w(\c_n|\c)=\sum_{n}w(\c|\c_n)P(\c_n),
\eeq
where, by hypothesis,
\beq
P(\c)\propto\e^{-E(\c)/T}.
\eeq
After division of both sides by the weight $P(\cal C)$,
eq.~(\ref{master0}) can be rewritten as
\beq\label{mast}
\sum_{n} w(\c_n|\c)-w(\c|\c_n)\e^{-\Delta E/T}=0,
\eeq
where the change in energy due to the flip reads
\beq\label{del}
\Delta E=E(\c_n)-E(\c)=2\s_n\,J(\s_{n-1}+\s_{n+1}).
\eeq
Detailed balance consists in equating each individual term appearing on both sides of the stationary master equation~(\ref{mast}), while global balance means that this equation is satisfied as a whole.
In other words~(\ref{mast}) expresses the global balance condition on the rates, which are the unknown quantities of the latter.
It leads to a set of constraints on the rates which should be satisfied in order for the stationary state to be Gibbsian with Hamiltonian~(\ref{hamilt}).
Before imposing any constraints the most general form of the rate function~$w(\s_n)$ takes eight values associated to the eight possible configurations of the group of spins $\{\s_n;\s_{n-1}\s_{n+1}\}$ involved in the flipping of the spin $\s_n$, and
thus depends a priori on eight parameters.
The constraint equations as well as additional symmetry conditions lower this number, as illustrated by what follows.

As shown in~\cite{gb2009}, eq.~(\ref{mast}) imposes the set of constraints
\beqa\label{taux0}
\w_{++}=\e^{4 K}\, w_{++}\nonumber\\
\w_{--}=\e^{-4 K}\,w_{--}\nonumber\\
\w_{+-}+\w_{-+}=w_{+-}+w_{-+}.
\eeqa
The compact notations
$w_{\s\s'}$ stands for $w(\s_n=+1;\s_{n-1}=\s,\s_{n+1}=\s')$, and
$\w_{\s\s'}$ for $w(\s_n=-1;\s_{n-1}=\s,\s_{n+1}=\s')$.
Note that the first two equations are detailed balance conditions.
These constraints lower the number of independent parameters to five.
Requiring that the $+$ and $-$ spins have symmetric roles, that is imposing the spin symmetry condition
\beq\label{spinsym}
w_{\s\s'}=\w_{-\s-\s'},
\eeq
introduces four additional constraints but reduces the set of previous constraints~(\ref{taux0}) to a single one,
\beq\label{taux1}
w_{--}=\e^{4 K}\, w_{++},
\eeq
which is precisely the detailed balance condition~(\ref{eq:db}).
The most general expression of the rate function compatible with the constraints~(\ref{spinsym}) and (\ref{taux1}), i.e., satisfying the global balance condition~(\ref{mast}), finally reads
\beqa\label{gener}
w(\s_n)&=&
\frac{1}{2}\alpha\big[1+\delta\,\s_{n-1}\s_{n+1}+\epsilon\,\s_{n-1}\s_{n}\nonumber
\\&-&(\gamma(1+\delta)+\epsilon)\s_{n}\s_{n+1}
\big],
\eeqa
depending now on the three independent parameters, $\alpha$, $\delta$, $\epsilon$,
and where the parameter $\gamma$ is given by~(\ref{gam})~\cite{gb2009}.
This form of the rate corresponds to the generic situation where the flipping spin is unequally influenced by the left and right spins.
This is the central result of this section.

From this general form, three particular cases emerge.

\begin{enumerate}
\item
The Glauber rate~(\ref{eq:glau}) is recovered if one imposes on~(\ref{gener}) that the left and right spins have an equal influence on the central spin. 
This right-left symmetry therefore requires that
\beq\label{sym}
w_{+-}=w_{-+},
\eeq
a constraint which itself fixes one of the remaining unknown parameters.
Carried in~(\ref{gener}) it yields $2 \epsilon+\gamma(1+\delta)=0$, hence
\beq
w(\s_n)=
\frac{1}{2}\alpha\big[1+\delta\s_{n-1}\s_{n+1}
-\frac{\gamma}{2}(1+\delta)\s_{n}(\s_{n-1}+\s_{n+1})\big].
\eeq
This expression, which now depends on two parameters, is the most general form of the rates for symmetric dynamics.
This form, originally proposed by Glauber (see Appendix of ref.~\cite{glau}), satisfies the detailed balance conditions~(\ref{taux1}) and (\ref{sym}).
The expression~(\ref{eq:glau}) is recovered by fixing $\delta=0$~\cite{glau}.

Note that in order to derive the form~(\ref{eq:glau}) we followed a different path than that chosen in~\cite{glau}. 
We first imposed the global balance condition then symmetry relations, while in~\cite{glau}
the spin and spatial symmetry conditions are already encoded in~(\ref{eq:glau}), then
the detailed balance condition~(\ref{taux1}) (i.e.,~(\ref{eq:db})) is imposed, leading to~(\ref{gam}), as recalled in the Introduction.

\item
The form of the rate~(\ref{direct}) used in the present work is obtained from~(\ref{gener}) with the particular choice of parameter $\delta=0$, and using the parameterization $\epsilon=-\gamma\, p$.

\item
It is interesting to note that the form obtained from~(\ref{direct}) for the totally asymmetric dynamics $p=0$ (or $p=1$) is, up to the choice of time scale encoded in $\alpha$, the {\it unique} choice at our disposal for a rate leading to a Gibbsian stationary state, as we now recall~\cite{gb2009}.
Assume for instance that the
spin $\s_n$ only looks to the right (i.e., $p=0$), hence
that the rates only depend on the right neighbour of the flipping spin:
\beq\label{taudir}
 w_{++}=w_{-+},\quad w_{+-}=w_{--}.
\eeq
Carrying these conditions into (\ref{gener}), we obtain $\delta=\epsilon=0$,
and therefore~\cite{gb2009}
\beq\label{rate:direct}
w(\s_n)=
\frac{1}{2}\alpha\left[1
-\gamma\,\s_{n}\s_{n+1}
\right],
\eeq
which is uniquely defined, up to the global time scale $\alpha$.
Fixing for example this scale by the choice $\alpha=2\cosh 2 K$ enables to write the rate in the compact exponential form
\beq\label{ku1D}
w(\s_n)=\e^{-2 K\s_{n}\s_{n+1}}.
\eeq
The totally asymmetric dynamics is much more constrained than the partially asymmetric one.
The existence of a rate function fulfilling the requested constraints~(\ref{taux0}), (\ref{spinsym}), and~(\ref{taudir}) was therefore not guaranteed a priori~\cite{gb2009}.

\end{enumerate}

\subsection{Rates in the presence of a magnetic field}

In the presence of a magnetic field, corresponding to adding to the Hamiltonian~(\ref{hamilt}) a term $\delta E=-h\sum_{n}\s_{n}$, the methods of ref.~\cite{gb2009} lead to the following constraint equations, with $H=h/T$,
\beqa\label{tauxH}
\w_{++}=\e^{4 K+2H}\, w_{++}\label{db1}\nonumber\\
\w_{--}=\e^{-4 K+2H}\,w_{--}\label{db2}\nonumber\\
\w_{+-}+\w_{-+}=\e^{2H}(w_{+-}+w_{-+}),
\eeqa
which are simple generalizations of~(\ref{taux0}).
The two first equations are conditions for detailed balance.
In the presence of a magnetic field we cannot impose the condition of spin symmetry~(\ref{spinsym}).
The most general expression of the rate function, corresponding to the general case of partially asymmetric dynamics, therefore depends on five parameters. 
This expression, that we do not write here, is the generalization of~(\ref{gener}) for rates depending on a magnetic field.
As above, imposing the additional constraints originating from particular choices of the symmetry of the dynamics lowers the number of independent parameters.
We consider successively the cases of symmetric, totally asymmetric and finally partially asymmetric dynamics.

\begin{enumerate}
\item
For symmetric dynamics, the additional constraints read
\beq\label{symgen}
w_{+-}=w_{-+},\qquad \w_{+-}=\w_{-+}.
\eeq
They lower the number of independent parameters to three and reduce the third equation of~(\ref{tauxH}) to the third detailed balance condition.
Instead of writing the associated general form for this case we give two simpler forms, both fully compatible with the constraint equations~(\ref{tauxH}) and (\ref{symgen}),
i.e., with detailed balance, as follows.
The first one,
\beq\label{glauH}
w(\s_n)=\frac{1}{2}\alpha\big[1-\frac{\gamma}{2}\s_n(\s_{n-1}+\s_{n+1})\big]
(1-\kappa \s_n),
\eeq
where $\alpha$ fixes the time scale, and $\kappa=\tanh H$,
was that proposed in~\cite{glau}.
A second simple expression of the rate, used in~\cite{gl2000}, is
\beq\label{glrate}
w(\s_n)=\frac{1}{2}\alpha\big[1-\s_n\tanh(K(\s_{n-1}+\s_{n+1})+H)\big].
\eeq

\item
For the totally asymmetric case, where for instance the flipping spin is only influenced by its right neigbour, the additional constraints are
\beq
 w_{++}=w_{-+},\quad w_{+-}=w_{--}, \quad \w_{++}=\w_{-+},\quad \w_{+-}=\w_{--}.
\eeq
The number of independent parameters is lowered to one, the time scale.
Up to this scale, the expression of the flipping rate is therefore once again {\it uniquely} determined.
It reads
\beq\label{ratedirect}
w(\s_n)
=\frac{1}{2}\alpha\left(1-\gamma\s_n\s_{n+1}\right)
(1-\kappa \s_n).
\eeq

\item
Coming back to the partially asymmetric case, we shall hereafter make the following choice of rate, inspired by the expressions~(\ref{glauH}) and~(\ref{ratedirect}), and interpolating between them,
\beq\label{rateasym}
w(\s_n)=\frac{1}{2}\alpha\left[1-\gamma\s_n(p\s_{n-1}+(1-p)\s_{n+1})\right]
(1-\kappa \s_n).
\eeq
This form of the rate is compatible with the three constraints~(\ref{tauxH}).
In particular it leads to~(\ref{ratedirect}) for $p=0$, as it should.

Let us mention that the natural generalization of~(\ref{glrate}), where the local field $K(\s_{n-1}+\s_{n+1})$ would be replaced by $2K(p\s_{n-1}+(1-p)\s_{n+1})$, does not lead to a rate function compatible with the constraints~(\ref{tauxH}), as long as $H\ne0$.

\end{enumerate}

Consider finally the situation where the system is subjected to a space-dependent arbitrary magnetic field $h_n$.
This corresponds to adding a perturbation of the form $\delta E=-\sum_n h_n\s_n$ to the ferromagnetic Hamiltonian~(\ref{hamilt}).
For symmetric dynamics it is easily seen that extending the rate functions~(\ref{glauH}) and (\ref{glrate}) by the replacement of the uniform field $h$ by the space-dependent field $h_n$ is compatible with detailed balance.
In contrast, in the presence of a space-dependent field, there is no simple form of the rate for asymmetric dynamics satisfying the global balance condition.
This will not prevent the possibility of defining a response function, which only requires to impose an infinitesimal field to the system, as will be seen later in this work.

\section{Biased magnetization}

The first quantity we shall want to investigate is the magnetization of the chain $M_n(t)=\langle \s_n(t)\rangle$. 
The Green function of the equations obeyed by the magnetization will appear as the cornerstone of much of the analysis of the next sections.

It is an easy matter to show that $M_n(t)$ obeys the evolution equations~\cite{glau}
\beq\label{eq:mag0}
\frac{\d \langle \s_n\rangle}{\d t}=-2\langle\s_n \,w(\s_n)\rangle,
\eeq
which yields, using~(\ref{direct}), and fixing the value of the parameter $\alpha$ to 1,
\beq\label{eq:mag}
\frac{\d M_n(t)}{\d t}=-M_n(t)+\gamma(p \,M_{n-1}(t)+(1-p)\,M_{n+1}(t)).
\eeq
With the choice of rate~(\ref{direct}), these equations are still linear, as was the case for symmetric dynamics~\cite{glau}.
They are thus easily solved by means of Laplace and Fourier transforms. 
Acting on the generic function $f_n(t)$, these transforms are defined as usual, as
\beq
f_{n}^{{\rm L}}(u)=\int_{0}^{\infty }\d t\, f_{n}(t)\,\e^{-u t},
\eeq
for the temporal Laplace transform, and as
\beq
f^{{\rm F}}(q,t)=\sum_{n}f_{n}(t)\,\e^{-{\i}nq},
\eeq
for the spatial Fourier transform with respect to the discrete space position $n$.
Combining the two transforms we shall also make use of the double Fourier-Laplace transform 
denoted by $f^{{\rm F}{\rm L}}(q,u)$.
Hereafter we shall consider an infinite system.

Using these integral transforms, (\ref{eq:mag}) becomes 
\beq
\frac{\d M^{{\rm F}}(q,t)}{\d t}=(\varphi(q)-1)M^{{\rm F}}(q,t),
\eeq
or 
\beq\label{eq:magfl}
u\,M^{{\rm F}{\rm L}}(q,u)=(\varphi(q)-1)M^{{\rm F}{\rm L}}(q,u)+M^{{\rm F%
}}(q,t=0), 
\eeq
yielding 
\beq\label{mfl}
M^{{\rm F}{\rm L}}(q,u)=\frac{M^{{\rm F}}(q,t=0)}{u+1-\varphi(q)},
\eeq
where the complex function $\varphi(q)$ has the alternate expressions
\beqa\label{phi}
\varphi(q)&=&\gamma\left( p\,\e^{-\i q}+(1-p)\,\e^{\i q}\right)
=\gamma(\cos q-\i V\sin q ),\label{phi2}\nonumber\\
&=&\gamma\sqrt{1-V^2}\cos(q-q_0),\qquad (\tanh\i q_0=V).
\eeqa

The Green function $G_{n}(t)$ is the solution of (\ref{eq:mag}) corresponding to
the locally magnetized initial condition where the spin at the origin is
pointing upwards, i.e., $\s _{0}(t=0)=+1$, but the configuration is
otherwise totally random. 
We thus have $G_{n}(t=0)=\delta _{n,0}$, i.e., 
$G^{{\rm F}}(q,t=0)=1$. 
Eq.~(\ref{mfl}) reads 
\beq\label{mfl1}
G^{{\rm F}{\rm L}}(q,u)=\frac{1}{u+1-\varphi(q)}, 
\eeq
hence 
\beq
G^{{\rm F}}(q,t)=\e^{-t(1-\varphi(q))},
\eeq
or
\beq\label{green}
G_{n}(t)=\e^{-t}\int_{0}^{2\pi }\frac{\d q}{2\pi }\,\e^{
t\varphi( q)+{\rm i}nq}.
\eeq
The expression of $G_n(t)$ in Laplace space is explicit and reads
\beqa\label{glap}
G_n^{\rm L}(u)=\int_{0}^{2\pi }\frac{\d q}{2\pi }\,\frac{\e^{\i nq}}{u+1-\varphi(q)}\nonumber
\\
=
\frac{\left((1+V)/(1-V)\right )^{n/2}}{\sqrt{(u+1)^2-\gamma^2(1-V^2)}}\left(\frac{u+1-\sqrt{(u+1)^2-\gamma^2(1-V^2)}}{\gamma
\sqrt{1-V^2}}\right)^{|n|}.\nonumber
\\
\eeqa
For symmetric Glauber dynamics with $V=0$, (\ref{green}) yields 
\beq\label{greenGlau}
G^{\mathrm{ sym}}_{n}(t)=\e^{-t}\int_{0}^{2\pi }\frac{\d q}{2\pi }\,\e^{\gamma
t\cos q+{\rm i}nq}=\e^{-t}I_{n}(\gamma t),
\eeq
where $I_{n}$ denotes the modified Bessel function of order $n$.
A generalization of this result to the asymmetric case is obtained as follows.
Consider the generating function
\beq\label{fctngene}
F(x,t)=\sum_{n=-\infty}^{\infty}G_{n}(t) x^n.
\eeq
Its explicit expression is 
\beq\label{fxt:asym}
F(x,t)=\e^{-t}\e^{\gamma t(px+(1-p)\frac{1}{x})},
\eeq
as can be seen by carrying~(\ref{fctngene}) into~(\ref{eq:mag}) and solving the resulting differential equation for $F(x,t)$.
Using the fact that
\beq
\e^{\frac{1}{2}y(z+\frac{1}{z})}=\sum_{n=-\infty}^{\infty}z^n\,I_n(y),
\eeq 
we obtain by identification 
\beq
F(x,t)=\e^{-t}\sum_{n=-\infty}^{\infty}x^n\left(\frac{1+V}{1-V}\right )^{n/2} I_{n}(\gamma t\,\sqrt{1-V^2}),
\eeq
hence the exact result
\beq\label{greenexact}
G_n(t)=\e^{-t}\left(\frac{1+V}{1-V}\right )^{n/2} I_{n}(\gamma t\,\sqrt{1-V^2}),
\eeq
which can easily be shown to be equivalent to~(\ref{green}), using~(\ref{greenGlau}), or alternatively to~(\ref{glap}).
The Green function is thus invariant in the simultaneous sign changes of $n$ and $V$.
Conversely, for a fixed value of $V$, $G_n(t)$ is spatially asymmetric,
as demonstrated by the expression of the time-independent ratio
\beq
\frac{G_n(t)}{G_{-n}(t)}=\left(\frac{1+V}{1-V}\right )^{n}.
\eeq
We thus define a length scale associated to this asymmetry by
\beq
\e^{\pm1/\xi_{\mathrm{ asym}}}=\sqrt{\frac{1+V}{1-V}},
\eeq
where the sign $\pm$ corresponds respectively to the cases $V>0$ and $V<0$.
This length, even in $V$, diverges for $V=0$ as $\xi_\mathrm{{asym}}\approx\pm 1/V$ and vanishes for $V=\pm1$\footnote{All throughout the present work the following notation is used for asymptotic equivalence: $f(x)\approx g(x)$ if $f(x)/g(x)\to1$.
The sign $\sim$ is used for asymptotic dominance by an exponential as in~(\ref{eq:decay}), or for indicating that two variables are of the same order of magnitude as in the example $\tau\sim\sqrt{s}$ encountered later on in the text. }.

Coming back to the general solution of (\ref{eq:mag}), its integral transform~(\ref
{mfl}) yields, in direct space, the discrete spatial convolution 
\beq\label{eq:conv}
M_{n}(t)=M_{n}(0)*G_{n}(t)=\sum_{m}M_{n-m}(0)G_m(t).
\eeq
In particular, for a uniformly magnetized system, with $M_n(t=0)=M$, we obtain
\beq\label{relax}
\frac{M_n(t)}{M}=\sum_n G_n(t)=G^\F(q=0,t)=\e^{-t/\tau_{\eq}},
\eeq
at any finite temperature, where the relaxation time $\tau_{\eq}$
is given by 
\beq\label{taueq}
\tau _{{\eq}}=\frac{1}{1-\gamma }.
\eeq
The result obtained in~(\ref{relax}) is seen to be independent of the value of the drift velocity $V$, the reason being that at zero Fourier momentum $\varphi(q=0)=\gamma$ no longer depends on $V$.
\begin{figure}
\begin{center}
\includegraphics[angle=0,width=.9\linewidth]{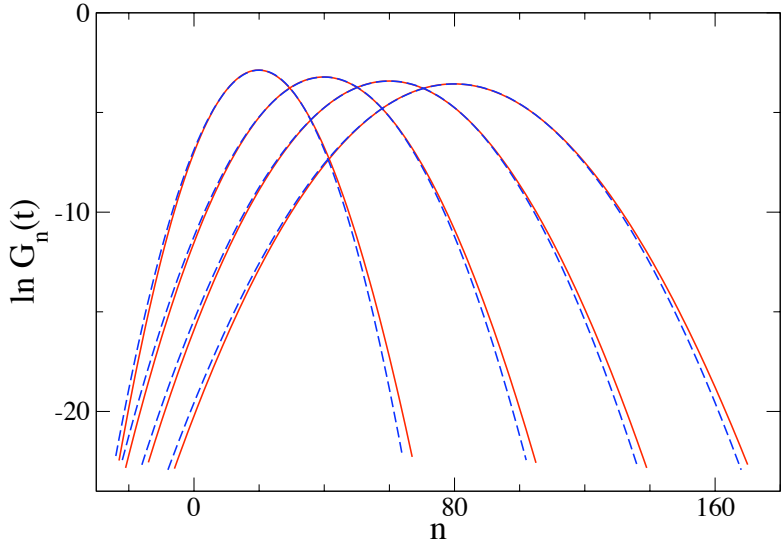}
\caption{\label{fig:green}
Green function at zero temperature, for $V=0.4$, at successive times $t=50,100, 150, 200$ (from left to right).
The locations of the maxima are at $n=V t$.
Continuous lines: exact result~(\ref{greenexact}).
Dashed lines: asymptotic estimate~(\ref{greenapprox}).
}
\end{center}
\end{figure}

At low temperature $\tau _{{\eq}}$ diverges exponentially fast according to 
$\tau _{{\eq}}\approx \e^{4 K}/2$.
The equilibrium correlation length of the Ising chain~\cite{baxter}, given by
\beq
\e^{-1/\xi_{\eq}}=\tanh K,
\label{xitau}
\eeq
also diverges exponentially fast as $\xi_\eq\approx \e^{2 K}/2$,
hence we have the scaling law
\beq
\tau_\eq\approx 2\xi_\eq^2,
\label{tauxi}
\eeq
corresponding to a dynamical critical exponent $z=2$,
and reflecting the diffusive nature of domain growth at zero temperature, even in the presence of a non zero drift velocity $V$.

The continuum limit estimate of $G_n(t)$ is obtained from~(\ref{green}), by using~(\ref{phi2}) and expanding the trigonometric functions around $q=0$:
\beqa
G_n(t)&\approx&\e^{-t}
\int_{0}^{2\pi }\frac{\d q}{2\pi }\,\e^{\gamma
t(1-q^2/2)+\i q(n-\gamma V t)},\nonumber
\\
&\approx&
\frac{1}{\sqrt{2\pi\gamma t}}\,\e^{-t(1-\gamma)}\e^{-\frac{(n-\gamma V t)^2}{2\gamma t}}.\label{greenapprox}
\eeqa
At any positive temperature this is an exponentially damped Gaussian centered at $n= \gamma V t$. 
At zero temperature this expression obeys the diffusion equation with drift
\beq\label{diffu}
\frac{\partial G_n(t)}{\partial t}=\frac{1}{2}\frac{\partial^2 G_n(t)}{\partial n^2}
-V\frac{\partial G_n(t)}{\partial n}.
\eeq
This estimate gives a faithful account of the bulk of the Green function but not of its tails, as depicted in Figure~\ref{fig:green}.
It is valid when time $t$ and space index $n$ are large and for small velocity $V$, i.e., for large length scale $\xi_{\mathrm{ asym}}$.

A far more accurate estimate of the Green function~(\ref{greenexact}) is obtained by performing the saddle point method on the integral representation of the Bessel function $I_n(t)$ given in~(\ref{greenGlau}), yielding
\beq\label{besselcol}
I_n(t)\approx\frac{1}{\sqrt{2\pi}}\frac{\e^{\sqrt{n^2+t^2}}}{(n^2+t^2)^{1/4}}
\left(\frac{t}{n+\sqrt{n^2+t^2}}\right)^n
.
\eeq
This estimate is valid when time $t$ and space $n$ are simultaneously large.
The Green function derived from~(\ref{greenexact}) and~(\ref{besselcol}) is undistinguishable from its exact value in Figure~\ref{fig:green}.
It is maximal for $n=\gamma V t$ at $t$ fixed, and, expanded around this maximum, yields back~(\ref{greenapprox}).
From the latter estimate~(\ref{greenapprox}) one obtains the value of $t$ where the Green function is maximal at $n$ fixed and large enough:
\beq\label{tstar}
t\approx\frac{n}{V}-\frac{1}{2V^2},
\eeq
if $\gamma=1$, and
\beq
t\approx\frac{1}{\sqrt{2-2\gamma+V^2\gamma}}
\left(\frac{n}{\sqrt{\gamma}}-\frac{1}{2\sqrt{2-2\gamma+V^2\gamma}}\right),
\eeq
in the general case.
At large time the asymptotic behaviour of the Green function is dominated by the exponential decay
\beq\label{eq:decay}
G_n(t)\sim\e^{-\alpha_{G}t},
\eeq
with relaxation rate (or inverse characteristic time)
\beq\label{alfaG}
\alpha_G=1-\gamma\sqrt{1-V^2},
\eeq
as can be obtained from~(\ref{greenexact}) and~(\ref{besselcol}), or from~(\ref{glap}).
We therefore confirm that the Gaussian approximation~(\ref{greenapprox}) is only valid for small $V$ since it predicts $\alpha_G=1-\gamma(1-V^2/2)$.

The case of the totally asymmetric dynamics, where $V=\pm1$,
deserves a separate treatment.
For example, for $V=1$, the Green function vanishes for $n<0$, while for $n\ge0$,
\beq\label{v1}
G_n(t)=\frac{\e^{-t}(\gamma t)^n}{n!}
\approx \frac{\e^{n-t}(\gamma t)^n }{\sqrt{2\pi }\,n^{n+1/2}}.
\eeq
This result can be obtained from the general form of the generating function~(\ref{fxt:asym}), yielding
\beq
F(x,t)=\e^{-t(1-\gamma x)}.
\eeq
It can also be obtained as the limit when $V\to1$ of~(\ref{greenexact}) using the expansion $I_n(x)\approx (x/2)^n/n!$ for $x\to0$.
Yet another means of obtaining the same result is by noting that, from~(\ref{glap}), we have
\beq
G_n^{\rm L}(u)=\frac{\gamma^n}{(u+1)^{n+1}}.
\eeq
Similarly, for $V=-1$, the Green function vanishes for $n>0$, while for $n\le0$,
\beq
G_{n}(t)=\frac{\e^{-t}(\gamma t)^{-n}}{(-n)!}.
\eeq

\section{Equal-time correlation function}

The next quantity we shall want to study is the two-spin equal-time correlation function
\beq
C(i,j,t)=\left\langle \s _{i}(t)\s _{j}(t)\right\rangle .
\eeq
For a random initial condition, averaging over all the
possible initial spin configurations and using the fact that the invariance under spatial
translations along the chain is preserved by the dynamics, the correlation
function only depends on the distance $|j-i|$ between the spins.
We denote it by
\beq
C(i,j,t)=C_{n}(t),
\eeq
with $n=j-i$.
In particular 
\beq\label{cond0}
C_{0}(t)=1. 
\eeq
Again, it is easy to derive the following evolution equation~\cite{glau}
\beq
\frac{\d \langle \s_i\s_j\rangle}{\d t}=-2\langle\s_i\s_j \,(w(\s_i)+w(\s_j)\rangle,
\eeq
for $C_n(t)$, which yields a set of coupled linear differential
equations of the form, 
\beq\label{eq:corr}
\frac{\d C_{n}(t)}{\d t}=-2C_{n}(t)+\gamma
(C_{n-1}(t)+C_{n+1}(t))\qquad (n\ne 0), \label{cdot}
\eeq
together with the condition~(\ref{cond0}), and the initial value $%
C_{n}(t=0)=\delta _{n,0}$.
These equations are insensitive to the presence of the bias $p$ as a consequence of the invariance of
the correlation function $\left\langle \s_{i}(t)\s_{j}(t)\right\rangle$ under the exchange of the indices $i$ and $j$.

This result is consistent with the conclusions of~\cite{gunter} where it was found that in the $A+A\to0$ system with driven diffusion, on which the model considered in the present work can be mapped at zero temperature, the local density fluctuations do not depend on the driving, for translationally invariant initial states. 
The latter are multispin generalizations of the equal-time correlation function studied here.

The method given in~\cite{gl2000} for the solution of~(\ref{cdot}) is now briefly recalled.
We complete~(\ref{cdot}) by the corresponding equation for $n=0$,
with a time-dependent source $v(t)$ in the right-hand side,
to be determined in such a way that the condition~(\ref{cond0}) be fulfilled.
In other words we consider the equations
\beq
\frac{\d C_n(t)}{\d t}=-2C_n(t)+\gamma(C_{n-1}(t)+C_{n+1}(t))+v(t)\delta_{n,0}.
\label{cdot0}
\eeq
The solution of this equation, 
together with the condition~(\ref{cond0}), and the initial value 
$C_{n}(t=0)=\delta _{n,0}$ is given in Fourier-Laplace space by~\cite{gl2000}
\beq\label{cfl}
C^{{\rm F}{\rm L}}(q,u)=\frac{v^{\rm L}(u)+1}{u+2-2\gamma\cos q}
=\frac{\sqrt{(u+2)^{2}-4\gamma ^{2}}}{u(u+2-2\gamma
\cos q)}. 
\eeq
hence
\beq\label{cnl}
C_n^{\rm L}(u)
=\frac{1}{u}
\left (\frac{u+2-\sqrt{(u+2)^2-4\gamma^2}}{2\gamma}\right )^{|n|}.
\eeq
We used the fact that~\cite{gl2000}
\beq\label{vlap}
v^{\rm L}(u)=\frac{1}{u}\sqrt{(u+2)^2-4\gamma^2}-1,
\eeq
the inverse of which is related to the correlation one site apart $C_1$ by
\beq\label{vt}
v(t)=2(1-\gamma C_1(t)).
\eeq
Eq.~(\ref{cnl}) yields limiting forms in the following two regimes of interest.
\begin{enumerate}

\item 

At finite temperature, for long enough times $(t\gg \tau _{{\eq}})$%
, the correlation function $C_{n}(t)$ converges to its equilibrium value $%
C_{n,{\eq}}$. 
We have 
\begin{equation}
C_{{\eq}}^{{\rm F}}(q)=\frac{\sqrt{1-\gamma ^{2}}}{1-\gamma \cos q},
\eeq
recovering thus the well-known expression 
$C_{n,{\eq}}=\e^{-\mu |n|}$~\cite{baxter}, where
\beq
\e^{-\mu}=\e^{-1/\xi_{\eq}}=\tanh K=\frac{1-\sqrt{1-\gamma^2}}{\gamma}=C_{1,{\eq}}.
\eeq

\item 
At zero temperature, and more generally in the aging regime 
$(t\ll \tau _{{\eq}})$, we obtain
\begin{equation}\label{cna}
C_{n}(t)\approx \,\mathop{\rm erfc}\left( \frac{|n|}{2\sqrt{t}}\right) ,
\end{equation}
where the error function $\mathop{\rm erf}z$ and the complementary error
function $\mathop{\rm erfc}z=1-\mathop{\rm erf}z$ are defined as 
\begin{equation}
\mathop{\rm erf}z=\frac{2}{\sqrt{\pi }}\int_{0}^{z}\d u\,\e^{-u^{2}}
,\quad \mathop{\rm erfc}z=\frac{2}{\sqrt{\pi }}%
\int_{z}^{\infty }\d u\,\e^{-u^{2}}.
\end{equation}
\end{enumerate}

\section{Two-time correlation function}
We now consider the two-time correlation function
\beq
C(i,j,s,t)=\left\langle \s_{i}(s)\s_{j}(t)\right\rangle ,
\eeq
where $s<t$  is the waiting time and $t=s+\tau$ the observation time. 
For a spatially homogeneous initial condition, as considered hereafter, we denote the correlation function by 
\beq
C(i,j,s,t)=C_{n}(s,t),
\eeq
with $n=j-i$.
Using the same methods as above, the two-time correlation function is readily shown to obey the coupled linear partial differential equations 
\beq\label{eq:cst}
\frac{\partial C_n(s,t)}{\partial t}=-C_n(s,t)+\gamma
(p\,C_{n-1}(s,t)+(1-p)\,C_{n+1}(s,t)), 
\eeq
for $t>{s}$, together with the initial condition 
\beq\label{cnss}
C_{n}(s,t={s})=C_{n}({s}).
\eeq
where the right-hand side is the equal-time correlation function, given in
Fourier-Laplace space by~(\ref{cfl}). 

\subsection{Solution of equation~(\ref{eq:cst})}
The first argument $s$ plays the role of a parameter in~(\ref{eq:cst}),
hence these equations are formally identical to~(\ref{eq:mag}),
with the initial condition~(\ref{cnss}) playing the role of the initial condition $M_n(t=0)$ for the latter.
The solution of~(\ref{eq:cst}), seen as an evolution equation
in the $\tau$ variable, is therefore formally identical to~(\ref{eq:conv}), and reads
\beq\label{conv}
C_n(s,s+\tau)=C_n(s)*G_n(\tau)=\sum_m C_{n-m}(s)G_{m}(\tau).
\eeq

In order to write down this solution more explicitly,
we introduce the double
Laplace transform of the function $C_n(s,s+\tau)$,
where $\ps$ is conjugate to $s$, the transform being denoted by $\Ls$,
and $v$ is conjugate to $\tau$, the transform being denoted by $\L$:
\beq
C_n^{\Ls\L}(\ps,v)
=\int_0^\infty\int_0^\infty \d s\,\d\tau\, C_n(s,s+\tau)\,\e^{-(\ps s+v\tau)}.
\label{defll}
\eeq
With this definition, the solution~(\ref{conv}) of (\ref{eq:cst}) reads
\beq\label{cfll}
C^{\F\Ls\L}(q,\ps,v)=\frac{C^{\F\L}(q,\ps)}{v+1-\varphi(q)},
\eeq
i.e., using the expression~(\ref{cfl}) of the equal-time correlation function in Fourier-Laplace space,
\beq\label{cfll1}
C^{\F\Ls\L}(q,\ps,v)
=\frac{\sqrt{(\ps+2)^2-4\gamma^2}}{\ps(\ps+2-2\gamma\cos q)(v+1-\varphi(q))}.
\eeq

An alternate expression is obtained as follows.
Denoting by
\beq
a^{\Ls}(u)=\frac{1}{u}\sqrt{(u+2)^{2}-4\gamma^{2}}=v^{\Ls}+1,
\eeq
and noting that the Green function for symmetric dynamics, $G^{\mathrm{ sym}}_{n}(s)$ is given in Fourier-Laplace space by
\beq
(G^{\mathrm{ sym}})^{{\rm F}{\Ls}}(q,u)=\frac{1}{u+1-\gamma \cos q},
\eeq
we get an expression for $C_n(s)$ in the form of a temporal convolution on $s$,
\beq
C_n(s)=a(s)*G^{\mathrm{ sym}}_{n}(2 s),
\eeq
hence from~(\ref{conv}), 
\beq\label{eq:alt}
C_n(s,s+\tau)
=a(s)*\sum_m G^{\mathrm{ sym}}_{n-m}(2 s)G_{m}(\tau),
\eeq
where, again, the convolution is on the variable $s$.

\subsection{Stationary regime}

At stationarity, letting $s\to\infty$, we find, from~(\ref{conv}) or from (\ref{cfll1}),
\beq\label{cstat}
C_{n,{\stat}}(\tau)=\e^{-\mu|n|}*G_n(\tau)
=\sqrt{1-\gamma^2}\int_{0}^{2\pi}\frac{\d q}{2\pi}
\frac{\e^{-\tau(1-\varphi(q))+\i n q}}{1-\gamma\cos q}.
\eeq
For symmetric dynamics, this expression simplifies to
\beq\label{csym}
C^\mathrm{ sym}_{n,{\eq}}(\tau)=\sqrt{1-\gamma^2}\int_{\tau}^{\infty}\d u\,G^{\mathrm{ sym}}_{n}(u),
\eeq
which is a monotonically decreasing function.
In contrast, the correlation~(\ref{cstat}) is not monotonic in $\tau=t-s$, except for $n=0$, 
hence its derivative with respect to this time variable has no constant sign.

In order to avoid this complication we concentrate on the case $n=0$ and inspect the asymptotic behaviour of $C_{0,{\stat}}(\tau)$ at large $\tau$.
The two extreme cases of symmetric ($V=0$) or totally asymmetric dynamics ($V=1$) are simpler.
For the former we have $C^\mathrm{ sym}_{0,{\eq}}(\tau)\sim\e^{-\tau(1-\gamma)}$, as can be seen from~(\ref{csym}) and (\ref{alfaG}).
For the latter, using the expression~(\ref{v1}) for $G_n(\tau)$ to compute the convolution appearing in (\ref{cstat}), we obtain the exact expression of the correlation as
\beq\label{eq:corrstat}
C_{0,{\stat}}(\tau)=\e^{-\tau \sqrt{1-\gamma^2}}.
\eeq
The question is therefore how to interpolate between the two previous decay laws, for partially asymmetric dynamics ($0<V<1$)\footnote{We consider here the case $V>0$, the extension to $V<0$ corresponding to changing the sign of $V$ where appropriate.}.
The task is to compute~(\ref{cstat}), or alternatively its Laplace transform
\beq\label{cfllstat}
C_{0,\stat}^{\L}(q,v)
=\sqrt{1-\gamma^2}\int_{0}^{2\pi}\frac{\d q}{2\pi}
\frac{1}{(1-\gamma\cos q)(v+1-\varphi(q))}.
\eeq
The integral in this equation is, setting $a=1/\gamma$ and $b=(v+1)/\gamma$, of the form
\beq
I(a,b,V)=\int_{0}^{2\pi}\frac{\d q}{2\pi}
\frac{1}{(a-\cos q)(b-\cos q+\i V\sin q)}.
\eeq
It is equal to
\beq\label{Iabv}
\frac{b^2-a^2 y^2}
{\sqrt{a^2-1} \sqrt{b^2-y^2} \left[\sqrt{a^2-1}\left(b-a y^2\right)
- (a-b) \sqrt{b^2-y^2}\right]},
\eeq
where $y^2$ is a temporary compact notation for $1-V^2$.
The leading singularities in the variable $b$ of this expression lie at $b_1=\sqrt{1-V^2}$ and $b_2=a-V\sqrt{a^2-1}$ (see the Appendix).
They correspond to the asymptotic behaviours, at exponential order,
\beq\label{c:decay}
C_{0,{\stat}}(\tau)\sim\e^{-\alpha_{1,2}\tau}
\eeq
with
\beq\label{c:alfa}
\alpha_{1}=1-\gamma \sqrt{1-V^2},\qquad
\alpha_{2}=V\sqrt{1-\gamma^2}.
\eeq
The relaxation rate $\alpha_1$ is identical to the relaxation rate $\alpha_G$ given in~(\ref{alfaG}).

For $1>V>V_c=\sqrt{1-\gamma^2}$ the two singularities coexist,
hence the decay is governed by the smaller rate, $\alpha_2<\alpha_1$.
However for $V<V_c$ the only leading singularity is at $b=b_1$,
corresponding to the relaxation rate $\alpha_1$.
For any positive temperature, there is therefore a critical velocity $V_c$ where a transition between two behaviours for $C_{0,{\stat}}(\tau)$ occurs.
At $V_c$ the relaxation rates are equal.
These two behaviours interpolate between the case $V=0$, where the relaxation rate is $\alpha_1=1-\gamma$ (unique singularity at $b_1$), and the case $V=1$, where the relaxation rate is $\alpha_2=\sqrt{1-\gamma^2}$ (unique singularity at $b_2$), as seen previously.
The same results hold if the spatial index $n\ne0$.

The physical interpretation of this behaviour is obtained by comparing the respective magnitudes of the two length scales considered earlier, namely the asymmetry length scale $\xi_\mathrm{ asym}$ and the equilibrium correlation length $\xi_{\eq}$.
The condition of weak asymmetry, $V<V_c$, corresponds to the condition $\xi_{\eq}<\xi_\mathrm{ asym}$,
the case of strong asymmetry to the condition $\xi_{\eq}>\xi_\mathrm{ asym}$.
\subsection{Zero-temperature scaling regime}
\label{Czero}

\subsubsection*{General theory}
We can rely on the Appendix to obtain general results on the behaviour of the correlation function, restricting the study to $C_0(s,t)$. 
This quantity is given by~(\ref{app:c0}).

For symmetric dynamics, the only singularity of the integral $I(a,b,0)$ lies at $b=b_1=1$, that is $v=0$.
This corresponds to a power-law decay of the correlation at large time, a result which is well-known, and recovered in~(\ref{eq:cscal}) below.

For totally asymmetric dynamics, the correlation is asymptotically given by the product of the pure exponential $\e^{-\tau}$ by an increasing function of $\tau$, which diverges as $\e^{2\sqrt{s\tau}}$, and thus sub-leading compared to the former.

For partially asymmetric dynamics, the analysis given in the Appendix predicts an asymptotic exponential decay of the correlation function, at large $\tau$, with $s$ fixed, with relaxation rate $\alpha_1=1-\sqrt{1-V^2}$ in the whole range $0< V<1$.
This prediction is confirmed by the analysis that follows.

\subsubsection*{Continuum limit}
Using the continuum limit expression of the Green function~(\ref{greenapprox}) we can estimate the sum appearing in~(\ref{eq:alt}) as
\beqa
\sum_m G^{\mathrm{ sym}}_{n-m}(2 s)G_{m}(\tau)
&\approx&
\frac{1}{2\pi\gamma\sqrt{s\tau}}
\int \d m \,\e^{-\frac{(n-m)^2}{4\gamma s}}\e^{-\frac{(m-\gamma V \tau)^2}{2\gamma\tau}}
\nonumber\\
&=&\frac{1}{\sqrt{2\pi\gamma(2s+\tau)}}
\e^{-(1-\gamma)(2s+\tau)}\e^{-\frac{(n-\gamma V\tau)^2}{2\gamma(2s+\tau)}}.
\nonumber
\eeqa
In the zero-temperature scaling regime where
$a^{\rm L}(u)\approx2/\sqrt{u}$, hence $a(s)\approx 2/\sqrt{\pi s}$,
we obtain, performing the convolution with $a(s)$,
\beq\label{cstscal}
C_n(s,s+\tau)\approx\frac{\sqrt{2}}{\pi}\int_0^s
\d u \frac{1}{\sqrt{u(2 s+\tau-2u)}}\e^{-\frac{(n-V\tau)^2}{2(2s+\tau-2u)}}.
\eeq
The changes of variable, $u=(2s+\tau)/(2+z^2)$, then $z=1/v$, followed by some algebra, enable us to recast~(\ref{cstscal}) into
\beq\label{cstscal2}
C_n(s,s+\tau)\approx\frac{1}{\sqrt{2\pi}}\int_{\frac{(n-V\tau)^2}{2s+\tau}}^\infty 
\d u\,\frac{\e^{-u/2}}{\sqrt{u}}\mathop{\rm erf}\left(\sqrt{\frac{su}{\tau}}\right).
\eeq
The virtue of this second form is that its derivative with respect to the time variable $s$ is explicit, as will be shown below.
These two equivalent expressions simplify in the following cases.
\begin{enumerate}
\item
If $\tau=0$ we have
\beq
C_n(s,s)\approx\mathop{\rm erfc} \left(\frac{|n|}{2\sqrt{s}}\right),
\eeq
in agreement with~(\ref{cna}).

\item
For $n=V\tau$, then
\beq\label{eq:cscal}
C_{n=V\tau}(s,s+\tau)\approx\frac{2}{\pi}\arctan\sqrt{\frac{2s}{\tau}}.
\eeq
This is also the low-temperature scaling form of the two-time correlation function at coincidence ($n=0$), in the symmetric case ($V=0$)~\cite{cox,bray89}, as can be readily checked in the expression~(\ref{cstscal}), since the only dependence in $n$ and $V$ is through $n-V\tau$.

\item
If $\tau\ll s$, then
\beq
C_n(s,s+\tau)\approx\mathop{\rm erfc} \left(\frac{|n-V\tau|}{2\sqrt{s}}\right).
\eeq
In particular, for $n=0$, we get anomalous aging, in the sense of~\cite{luckmehta}.
The two time variables $\tau$ and $s$ should be scaled as $\tau\sim\sqrt{s}$, which gives  
the bulk of the correlation.

\item
Finally if $\tau\gg s$, using the fact that the integral in~(\ref{cstscal}) is dominated by its lower bound, we find
\beq
C_n(s,s+\tau)\approx \sqrt{\frac{2}{\pi}}\,
\e^{V^2s}
\mathop{\rm erf}(V\sqrt{s})\frac{\e^{-\frac{V^2\tau}{2}}}{V\sqrt{\tau}}
.
\eeq
The relaxation rate of the exponential decay at large $\tau$ is equal to $V^2/2$, which is the first term of the expansion of $\alpha_1$ in $V$, in agreement with the prediction done above.

\end{enumerate}

Note that the continuum limit zero-temperature expressions~(\ref{cstscal}) or~(\ref{cstscal2}) of $C_n(s,t)$ satisfy the diffusion equation with drift~(\ref{diffu}).

%
\subsection{Temporal derivatives of the two-time correlation function}

In preparation of the forthcoming discussion of the fluctuation-dissipation relation, we need know
the expressions of the derivatives of $C_n(s,t)$ with respect to one of the time variables $s$ or $t$, the other being kept fixed.

From~(\ref{conv}) we have
\beq
\frac{\partial C_n(s,s+\tau)}{\partial s}=\frac{\d C_n(s)}{\d s}
*G_n(\tau)+
C_n(s)*\frac{\partial G_n(\tau)}{\partial s},
\eeq
with, using (\ref{cdot0}),
\beqa
\frac{\d C_n(s)}{\d s}*G_n(\tau)&=&
-2C_n(s,t)+\gamma(C_{n-1}(s,t)+C_{n+1}(s,t))\nonumber\\
&+&v(s)G_n(\tau),
\eeqa
and, using (\ref{eq:mag}),
\beqa\label{dcs}
C_n(s)*\frac{\partial G_n(\tau)}{\partial s}&=&
C_n(s,t)-\gamma(p C_{n-1}(s,t)+(1-p)C_{n+1}(s,t))\nonumber\\
&=&-\frac{\partial C_n(s,s+\tau)}{\partial t}.
\eeqa
Finally
\beqa\label{eq:dcs}
\frac{\partial C_n(s,s+\tau)}{\partial s}&=&
-C_n(s,t)+\gamma((1-p) C_{n-1}(s,t)+pC_{n+1}(s,t))\nonumber\\
&+&v(s)G_n(\tau).
\eeqa
This derivative obeys the same evolution equation with respect to time $t$ as that obeyed by 
$M_n(t)$ and $C_n(s,t)$. 
Its initial condition at $t=s$ reads
\beqa
\frac{\partial C_n(s,s+\tau)}{\partial s}\vert_{t=s}&=&
-C_n(s)+\gamma((1-p) C_{n-1}(s)+pC_{n+1}(s))\nonumber\\
&+&v(s)\delta_{n,0}.
\eeqa
We also note that
\beqa\label{zz}
\frac{\partial C_n(s,s+\tau)}{\partial s}-\frac{\partial C_n(s,s+\tau)}{\partial t}
=v(s)G_n(\tau)\nonumber\\
+\gamma V (C_{n+1}(s,s+\tau)-C_{n-1}(s,s+\tau)).
\eeqa

\begin{figure}
\begin{center}
\includegraphics[angle=0,width=.9\linewidth]{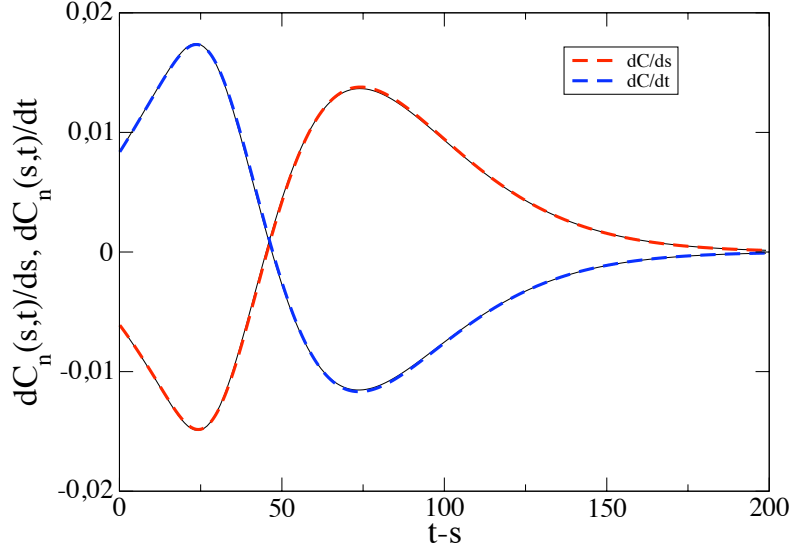}
\caption{\label{fig:Dcorrel}
Derivatives of $C_{n}(s,t)$ with respect to $s$ and $t$ in the zero-temperature scaling regime ($n=20$, $s=80$, $\gamma=1$, $V=0.4$).
The derivative with respect to $s$ ($t$) is first negative (positive), then positive (negative).
Continuous lines: asymptotic expressions~(\ref{dcsscal}) and (\ref{dctscal}). 
Dashed lines: numerical integration of the equations obeyed by the derivatives.
}
\end{center}
\end{figure}

In the zero-temperature scaling regime we have explicit expressions of these derivatives.
Using~(\ref{cstscal2}), we find
\beqa\label{dcsscal}
\frac{\partial C_n(s,s+\tau)}{\partial s}\approx\frac{\sqrt{2}}{\pi}\frac{s+\tau}{(2s+\tau)\sqrt{s\tau}}
\e^{-\frac{(n-V\tau)^2}{2\tau}}\nonumber\\
+\frac{1}{\sqrt{2\pi}}\frac{n-V(4s+3\tau)}{(2s+\tau)^{3/2}}\e^{-\frac{(n-V\tau)^2}{2(2s+\tau)}}
\mathop{\rm erf} \left[\frac{(n-V\tau)\sqrt{s}}{\sqrt{\tau(2s+\tau)}}\right],
\eeqa
and
\beqa\label{dctscal}
\frac{\partial C_n(s,s+\tau)}{\partial t}
\approx-\frac{\sqrt{2}}{\pi}\frac{\sqrt{s}}{(2s+\tau)\sqrt{\tau}}
\e^{-\frac{(n-V\tau)^2}{2\tau}}\nonumber\\
+\frac{1}{\sqrt{2\pi}}\frac{n+V(4s+\tau)}{(2s+\tau)^{3/2}}\e^{-\frac{(n-V\tau)^2}{2(2s+\tau)}}
\mathop{\rm erf} \left[\frac{(n-V\tau)\sqrt{s}}{\sqrt{\tau(2s+\tau)}}\right].
\eeqa
Hence, 
\beqa
\frac{\partial C_n(s,s+\tau)}{\partial s}-\frac{\partial C_n(s,s+\tau)}{\partial t}\approx
\frac{\sqrt{2}}{\pi}\,\frac{\e^{-\frac{(n-V\tau)^2}{2\tau}}
}{\sqrt{s\tau}}\nonumber\\
-\frac{4 V}{\sqrt{2\pi(2s+\tau)}}\e^{-\frac{(n-V\tau)^2}{2(2s+\tau)}}
\mathop{\rm erf} \left[\frac{(n-V\tau)\sqrt{s}}{\sqrt{\tau(2s+\tau)}}\right].
\eeqa

As depicted in Figure~\ref{fig:Dcorrel}, both derivatives have no constant sign, for $n$ fixed.
The location where these derivatives vanish, $n$ being fixed, can be found by the following analysis.
Let $s$ be large ($s\gg n, \tau$), and define the variable $\zeta=(n-V\tau)/\sqrt{2\tau}$.
Then from~(\ref{dcsscal}) and~(\ref{dctscal}) we have, for $\zeta$ small,
\beq
\e^{\zeta^2}\mathop{\rm erf}\zeta\approx\frac{2}{\sqrt{\pi}}\zeta\approx\frac{1}{\sqrt{2\pi V^2\tau}},
\eeq
showing that the two derivatives vanish for $\tau\approx n/V-1/2V^2$, at a fixed, large enough value of $n$.
This result is similar to~(\ref{tstar}).

\section{Response function and fluctuation-dissipation relation}

\subsection{Overview}

Just as the two-time correlation function characterizes the spontaneous fluctuations of a system at thermal equilibrium, the response function characterizes how the system gently driven out of equilibrium by a small external perturbation returns to its equilibrium state.
In the present case of a spin system, we define the response as the rate of change of the limiting magnetization induced by an infinitesimal field applied impulsively on the system.
For a time-reversible process, the fluctuation-dissipation theorem states that, at equilibrium,
the response and correlation functions obey the equality 
\beq\label{fdt}
R_{\eq}(\tau)=-\frac{1}{T}\frac{\d C_{\eq}(\tau)}{\d \tau},
\eeq
whose physical content is that spontaneous fluctuations and fluctuations induced by a small external perturbation have the same decay laws.

Attempts to generalize this theorem to nonequilibrium situations opened several directions of research. 
\begin{enumerate}
\item
The first possible generalization consists in trying to define a temperature for a nonequilibrium system.
A summary of early attempts and contributions in this direction is given in~\cite{ck1}.
A natural extension of the concept of a temperature was then proposed in~\cite{ck1,ck2,ck3}
for aging systems, such as glasses or systems exhibiting domain growth, as well as for systems gently driven out of equilibrium, through the introduction of 
the fluctuation-dissipation ratio $X(s,t)$ which is the ratio of the response by the temporal derivative of the correlation with respect to the waiting time $s$ (precise definitions are given in~(\ref{R}) and~(\ref{Xdef})).
This quantity can be interpreted as giving a measure of the distance to equilibrium of the system.
As such this definition has no further content.
But for the systems of interest considered in~\cite{ck1,ck2,ck3} $X(s,t)$ turns out to be a non trivial function of its two arguments, depending on the two times $s$ and $t$ through $C(s,t)$ only.
Different functional dependences of $X(s,t)$ on $C(s,t)$ correspond to different classes of out of equilibrium systems~\cite{ck1}.

\item
The second generalization consists in trying to obtain expressions of the response in terms of correlations characteristic of the non perturbed system~\cite{ck2,chate,fede,corb,diez,seif,maes,corb3}.
When applied to non time-reversible systems,
this investigation relies on the prior necessary question of how to define the response.

\item
A different viewpoint originates from the discovery that, at criticality,
for systems undergoing critical coarsening, the limit fluctuation-dissipation ratio $X_\infty$, defined in~(\ref{xinfty}),
is a dimensionless amplitude ratio, and therefore a novel universal quantity in nonequilibrium critical dynamics~\cite{gl2000,gl2000+,gl2002}.
This was demonstrated by the analysis of several ferromagnetic
spin systems after a quench from infinite temperature to their critical temperature, namely
the one-dimensional Glauber-Ising chain, the two-dimensional Ising model and the spherical model in any dimension~\cite{gl2000,gl2000+,gl2002}. 
Numerous further investigations confirmed the universal character of the critical fluctuation-dissipation ratio  in a variety of systems~\cite{calabrese,sollich2,sollich3,corb2, malte}.

In contrast, analytical and numerical studies indicate that the limit fluctuation-dissipation ratio vanishes throughout the low-temperature phase of coarsening systems~\cite{ck1,x,gl2002}.

\end{enumerate}

We shall be confronted with these three facets of the generalization of the fluctuation-dissipation theorem when examining the behaviour of the response for the directed Ising chain.
In the present situation, the asymmetry of the dynamics already implies the violation of the fluctuation-dissipation theorem at stationarity.

A preliminary question is how to define the response for such a non time-reversible system.
As mentioned earlier, the rate function~(\ref{rateasym}) no longer satisfies the global balance condition for the perturbed Hamiltonian if the applied magnetic field is space dependent.
We shall nevertheless keep the same functional form of the rate because the perturbation is infinitesimal, thus the system stays infinitely close to a stationary state satisfying this global balance condition.

\subsection{Definition of the response function} 
  
We define the response function in the same way as for symmetric dynamics~\cite{gl2000}.  
We suppose that the system is subjected to a small arbitrary magnetic field $h_n(t)$.
We define the dynamics of the model by the rate~(\ref{rateasym}),
where $\kappa$ is be replaced by $\kappa_n(t)=\tanh H_n(t)$, with $H_n(t)=h_n(t)/T$.
The magnetization $M_n(t)$ at time $t$ reads
\beq
M_n(t)=\mean{\s_n(t)}\approx\int_0^t\d u\sum_m R_{n-m}(u,t)H_m(u),
\eeq
to first order in the magnetic field $H_n(t)$, which defines the
dimensionless two-time response function $R_{n-m}(s,t)$, thus given by
the functional derivative
\beq\label{R}
R_{n-m}(s,t)=\left.\frac{\delta M_n(t)}{\delta H_m(s)}\right\vert_{\{H=0\}}.
\eeq

The evolution equation~(\ref{eq:mag0}) yields
the inhomogeneous differential equations for the magnetization
\beqa\
\frac{\d M_n(t)}{\d t}&=&-M_n(t)+\gamma(p \,M_{n-1}(t)+(1-p)\,M_{n+1}(t))\nonumber\\
&+&\kappa_n(t)(1-\gamma(pC(n-1,n,t)+(1-p)C(n,n+1,t))),
\eeqa
that is, to first order in the magnetic field, 
\beqa\label{eq:magH}
\frac{\d M_n(t)}{\d t}&=&-M_n(t)+\gamma(p \,M_{n-1}(t)+(1-p)\,M_{n+1}(t))
\nonumber\\
&+&H_n(t)(1-\gamma C_1(t)).
\eeqa
As a consequence, the two-time response function $R_n(s,t)$ itself
obeys the linear differential equations
\beq\label{eq:rst}
\frac{\partial R_n(s,t)}{\partial t}
=-R_n(s,t)+\gamma (p R_{n-1}(s,t)+(1-p)R_{n+1}(s,t)),
\eeq
for $t>s$, with the initial value
\beq
R_n(s,s)=w(s)\delta_{n,0},
\label{rnss}
\eeq
and with $w(s)=1-\gamma C_1(s)=v(s)/2$ (see~(\ref{vt})), where $C_1(s)$ is the equal-time correlation one site apart,
given in Laplace space by~(\ref{cnl}).
Equation~(\ref{eq:rst}), with its initial condition~(\ref{rnss}),
is formally identical to~(\ref{eq:cst}) with initial condition~(\ref{cnss}).
Hence its solution reads
\beq\label{eq:rnst}
R_n(s,s+\tau)=w(s) G_n(\tau).
\eeq

On view of~(\ref{zz}) and (\ref{eq:rnst}) the response function can be expressed in terms of the correlation function $C_n(s,t)$ and its derivatives as
\beqa\label{rc}
R_n(s,t)&=&\frac{1}{2}\left(\frac{\partial C_n(s,t)}{\partial s}-
\frac{\partial C_n(s,t)}{\partial t}\right)\nonumber\\
&-&\frac{\gamma V}{2}(C_{n+1}(s,t)-C_{n-1}(s,t)).
\eeqa
The response is thus expressed in terms of quantities of the non perturbed system,
as is already clear from~(\ref{eq:rnst}).
Generalized fluctuation-dissipation relations of this kind have been investigated previously~\cite{ck2,corb,diez,seif,maes}.
The form~(\ref{rc}) can be identified with
the general expression for the response function given in~\cite{corb}, by noting that
the term proportional to the drift velocity $V$ in~(\ref{rc}), which is naturally associated to the asymmetry due to the bias, is precisely equal to the expression named ``asymmetry'' in~\cite{corb}, 
\beq
A_{i,j}(s,t)=\frac{1}{2}[\langle \s_i(t)B_j(s)\rangle-\langle \s_j(s)B_i(t)\rangle],
\eeq
where $B_i(t)=-2\s_i w(\s_i)$ and $n=j-i$.

However~(\ref{rc}) does not hold in all generality.
Its validity relies on a particular choice made for the form of the rate function.
Let us take the simpler case $V=0$ to illustrate the point.
Taking the form~(\ref{glrate}) for the rate, as done in~\cite{gl2000}, leads to a different expression for $w(s)$, hence for $R_n$.
As found in~\cite{gl2000}, $w(s)=1-\gamma^2(1+C_2(s))/2$.
The corresponding response function reads
\beq\label{eq:modif}
R_n(s,t)=\frac{1}{2}\left(\frac{\partial C_n(s,t)}{\partial s}-
\frac{\partial C_n(s,t)}{\partial t}\right)-\frac{\gamma}{2}\frac{\d C_1(s)}{\d s}G_n(\tau).
\eeq
The reason for such a discrepancy lies in the choice made in~\cite{corb} for the rate in the presence of a field.
It requests the latter to be of the form 
\beq
w^{h}(\s_n)=w^{h=0}(\s_n)\Delta w(\s_n),
\eeq
where $\Delta w$ is a function of $\s_n$, with no dependence on the neighbouring spins, which is not the case for the form~(\ref{glrate}).
However, the additional term appearing in~(\ref{eq:modif}) vanishes identically at stationarity, and asymptotically in the zero-temperature scaling regime, leading in both cases to a unique definition of the response function.

\subsection{Fluctuation-dissipation relation in the presence of a uniform field}
As we noticed earlier (see~(\ref{relax})), summing upon the spatial index $n$ i.e., taking the zero Fourier component of quantities, wipes out the role of the asymmetry, since $\varphi(q=0)=\gamma$ does not depend on $V$.
In the present case summing the two sides of~(\ref{rc}) upon the spatial index $n$ 
suppresses the contribution coming from the second term, proportional to $V$, in the right side of the equation, yielding
\beq\label{eq:rf0}
R^{\rm F}(q=0,s,t)
=\frac{1}{2}\left(\frac{\partial C^{\rm F}(q=0,s,t)}{\partial s}-
\frac{\partial C^{\rm F}(q=0,s,t)}{\partial t}\right),
\eeq
with
\beq
\sum_{n}R_n(s,t)=R^{\rm F}(q=0,s,t)=w(s)\e^{-\tau/\tau_{\eq}},
\eeq
(see (\ref{relax})) and
\beq
\sum_{n}C_n(s,t)=C^{\rm F}(q=0,s,t)=\chi(s)\e^{-\tau/\tau_{\eq}}.
\eeq
The susceptibility
\beq
\chi(s)=\sum_nC_n(s)=C^\F(q=0,s ),
\label{cbc}
\eeq
given in Laplace space by (see~(\ref{cfl}))
\beq
\chi^{\rm L}(u)=\frac{1}{u}\sqrt{\frac{u+2+2\gamma}{u+2-2\gamma}}.
\eeq
measures the spatial range over which ferromagnetic order has propagated
at time $s$~\cite{gl2000}.
Hence $C^{\rm F}(q=0,s,t)$ and $R^{\rm F}(q=0,s,t)$  are independent of $V$, i.e., have the same forms as for symmetric dynamics.
In other words, under a uniform magnetic field the response function is insensitive to the asymmetry of the dynamics.%

\subsection{Fluctuation-dissipation relation in the stationary state}

\begin{figure}
\begin{center}
\includegraphics[angle=0,width=.9\linewidth]{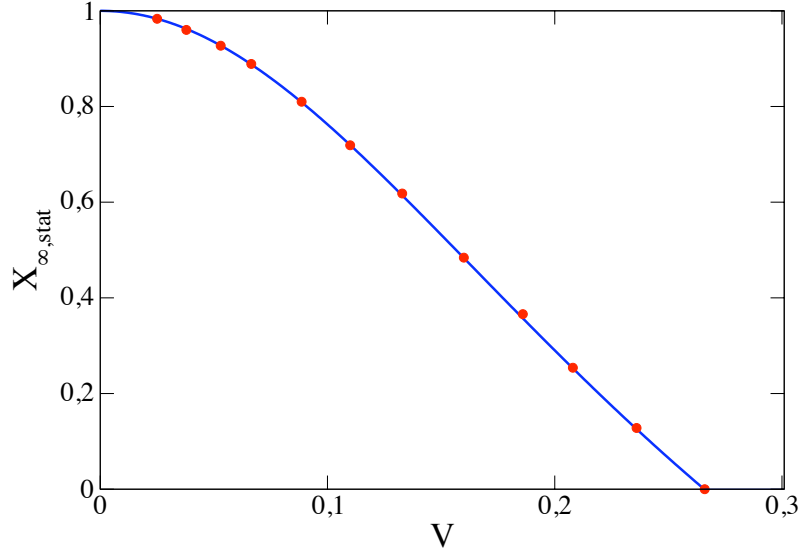}
\caption{\label{fig:X}
Stationary limit fluctuation-dissipation ratio versus velocity $V$, for $V<V_c=\sqrt{1-\gamma^2}=0.2658\ldots$ ($\gamma=\tanh 2$).
Continuous line: analytical prediction~(\ref{X_stat}); dots: numerical integration of the temporal equations for $R_0$ and $\d C_0/\d \tau$. 
For $V>V_c$, $X_{\infty,{\stat}}=0$.}
\end{center}
\end{figure}

An immediate consequence of~(\ref{eq:rf0}) is that the fluctuation-dissipation theorem holds in the stationary state for the Fourier transforms at zero momentum, i.e., when applying a uniform magnetic field, 
\beq
R_{\stat}^{\rm F}(q=0,\tau)=-\frac{\d C_{\stat}^{\rm F}(q=0,\tau)}{\d \tau}.
\eeq

However, in the general case, the fluctuation-dissipation theorem does not hold at stationarity.
This can be demonstrated either using~(\ref{rc}), which yields
\beq\label{fdstat}
R_{n,\stat}(\tau)=-\frac{\d C_{n,\stat}(\tau)}{\d \tau}
-\frac{\gamma V}{2}(C_{n+1,\stat}(\tau)-C_{n-1,\stat}(\tau)),
\eeq
the presence of the second term in the right side ruling out the theorem,
or alternatively, using~(\ref{eq:rnst}), yielding
\beq\label{rstat}
R_{n,{\stat}}(\tau)=\sqrt{1-\gamma^2}\,G_{n}(\tau),
\eeq
which cannot be identified with the time derivative of $C_{n,{\stat}}(\tau)$, given by~(\ref{cstat}).
Note that, if $n\ne0$, this violation is strong, because, as noticed earlier, the derivative of the correlation with respect to $\tau$ has no constant sign, for $n$ fixed, and therefore can in no way be identified to the response function, which is always positive.

\begin{figure}
\begin{center}
\includegraphics[angle=0,width=.9\linewidth]{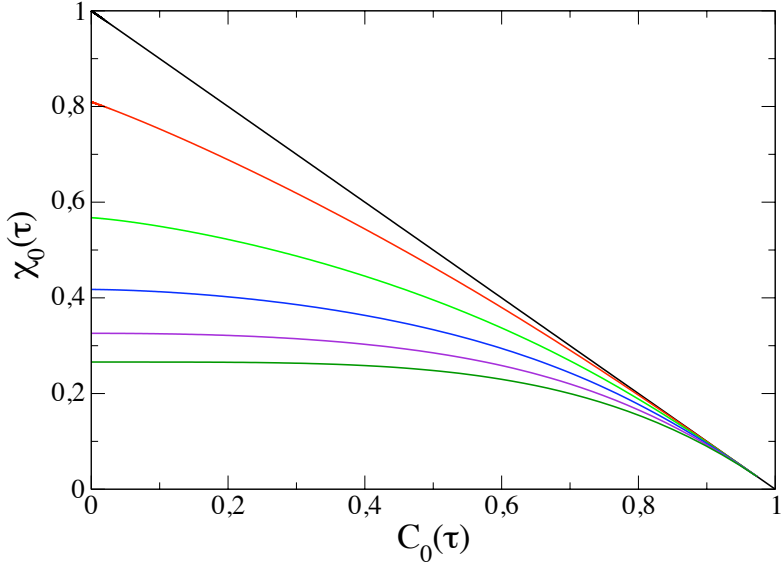}
\caption{\label{fig:chi}
Stationary integrated response function against stationary correlation.
From top to bottom, $V=0,0.2,0.4,\ldots, 1$ ($\gamma=\tanh 2$).
}
\end{center}
\end{figure}

We now investigate the relative asymptotic behaviours of $R_{0,{\stat}}(\tau)$ and $-\d C_{0,{\stat}}(\tau)/\d \tau$.
The study of the case $n\ne0$ leads to the same conclusions.
By~(\ref{besselcol}) we know the asymptotic decay law of $R_{0,{\stat}}$ at large $\tau$,
\beq
R_{0,{\stat}}(\tau)\approx\sqrt{1-\gamma^2}\,
\frac{\e^{-\tau(1-\gamma\sqrt{1-V^2})}}{\sqrt{2\pi\gamma\tau\sqrt{1-V^2}}},
\eeq
and the decay of $C_{0,{\stat}}(\tau)$ is given by~(\ref{c:decay}) and~(\ref{c:alfa}).

Therefore, if $V>V_c=\sqrt{1-\gamma^2}$, the stationary limit fluctuation-dissipation ratio defined as
\beq
X_{\infty,\stat}=\lim_{\tau\to\infty}\frac{R_{0,{\stat}}(\tau)}{-\d C_{0,\stat}(\tau)/\d \tau}
\eeq
vanishes, because the relaxation rate of the response
$\alpha_1=\alpha_G=1-\gamma\sqrt{1-V^2}$ is always larger than the relaxation rate of the correlation
$\alpha_2=V\sqrt{1-\gamma^2}$.
However for $V<V_c$ the relaxation rates of the two quantities are both equal to $\alpha_1$.
It turns out that the sub-leading corrections to the exponential decay are also the same, i.e., defining the amplitude $A_R$ and $A_{C'}$ by
\beqa
R_{0,{\stat}}(\tau)\approx A_R(\tau)\e^{-\tau(1-\gamma\sqrt{1-V^2})}\\
-\frac{\d C_{0,\stat}(\tau)}{\d \tau}\approx A_{C'}(\tau)\e^{-\tau(1-\gamma\sqrt{1-V^2})},
\eeqa
we have
\beqa\label{X_stat}
X_{\infty,\stat}&=&\lim_{\tau\to\infty}\frac{A_{R}(\tau)}{A_{C'}(\tau)}
=\lim_{v\to-\alpha_1}\frac{R_{0,\stat}^{\L}(v)}{v\, C_{0,\stat}^{\L}(v)}\nonumber\\
&=&\frac{1-\gamma/\sqrt{1-V^2}}{1-\gamma\sqrt{1-V^2}}.
\eeqa
The Laplace transforms $C_{0,\stat}^{\L}(v)$ and $R_{0,\stat}^{\L}(v)$ are respectively given by~(\ref{cfllstat}) and~(\ref{Iabv}), and by~(\ref{glap}), while $\alpha_1$ is defined in~(\ref{c:alfa}).

In summary, if $V<V_c$, the stationary fluctuation-dissipation ratio has a finite limit at large time because the correlation and response functions have the same asymptotic exponential decay.
For $V>V_c$ the correlation function decays more slowly and $X_{\infty,\stat}=0$.
The unexpected result obtained here is that this effect leads to a transition in the behaviour of the  fluctuation-dissipation ratio.
As mentioned earlier, this transition is due to the crossing of two length scales, namely the equilibrium correlation length $\xi_{\eq}$ and the asymmetry length scale $\xi_{\mathrm{ asym}}$.
The condition $V<V_c$ corresponds to the condition $\xi_{\eq}<\xi_\mathrm{ asym}$.
The effect of the asymmetry becomes dominant when $\xi_{\eq}>\xi_\mathrm{ asym}$.

Figure~\ref{fig:X} gives a comparison between this prediction and the data obtained by a numerical integration of the temporal equations for the correlation and response functions.
These data points were obtained by an extrapolation of a plot of the ratio $-R_{0,{\stat}}(\tau)/\d C_{0,{\stat}}(\tau)/\d \tau$ against $1/\tau$ for $V<V_c$, or against $1/\sqrt{\tau}$ when approaching the transition.

Another representation of the relation between correlation and response is obtained by using the integrated response function
\beq
\chi_{0}(s,t)=\int_{s}^{t}\d u\,R_{0}(u,t).
\eeq
In the stationary state we have $R_0(u,t)=R_{0,{\stat}}(t-u)$, hence
\beq
\chi_{0,{\stat}}(\tau)=
\int_{0}^{\tau}\d u\,R_{0,{\stat}}(u)=\chi_{\infty}-\int_{\tau}^{\infty}\d u\,R_{0,{\stat}}(u),
\eeq
with $\chi_{\infty}=\sqrt{1-\gamma^2}/\sqrt{1-\gamma^2(1-V^2)}$.
In particular, at equilibrium, for $V=0$, 
$
\chi_{0,{\eq}}(\tau)=1-C_{0,{\eq}}(\tau)
$.
Asymptotically
\beq
\chi_{0,{\stat}}(\tau)\approx\chi_{\infty}-\frac{A_{R}(\tau)}{\alpha_1}\e^{-\alpha_1\tau}.
\eeq
Hence the asymptotic ratio $X_{\infty,\stat}$ can be extracted from the slope at the origin of a plot of $\chi_{0,{\stat}}(\tau)$ against $C_{0,{\stat}}$.
Such a plot is depicted in figure~\ref{fig:chi}, for several values of the velocity $V$ and for a fixed value of the temperature.
The plot was obtained by a numerical integration of the equations satisfied by $\chi_n(s,t)$.

\subsection{Fluctuation-dissipation ratio in the zero-temperature scaling regime}

In the presence of the two time variables $s$ and $t=s+\tau$, we first define the asymptotic fluctuation-dissipation ratio~\cite{gl2000} as
\beq
X_{\mathrm{ as}}(s)=\lim_{\tau\to\infty}X_n(s,s+\tau)\equiv
\lim_{t\to\infty}X_n(s,t)
\eeq
then its limit at large waiting time $s$,
\beq\label{xinfty}
X_{\infty}=\lim_{s\to\infty}X_{\mathrm{ as}}(s)=\lim_{s\to\infty}\lim_{t\to\infty}X_n(s,t),
\eeq
where the spatially dependent fluctuation-dissipation ratio reads
\beq\label{Xdef}
X_n(s,t)=\frac{R_n(s,t)}{\partial C_n(s,t)/\partial s}.
\eeq

\subsubsection*{General theory}
The study of the decay of the correlation function, done in section~\ref{Czero}, leads to the following predictions.
For symmetric dynamics the limit fluctuation-dissipation ratio $X_\infty$ is finite, as is well known and as shown below.
For the totally asymmetric dynamics we have
$X_{\mathrm{ as}}(s)=0$ for any waiting time $s$ because the decay of the response is faster than that of the correlation function.
For partially asymmetric dynamics ($0<V<1$) we predict a finite asymptotic fluctuation-dissipation ratio $X_{\mathrm{ as}}(s)$ which decays to $0$ for $s\to\infty$. 
This last prediction is confirmed by the analysis of the continuum limit.

\subsubsection*{Continuum limit}

In the regime of large times and $V$ small enough such that~(\ref{greenapprox}) is a valid approximation of the Green function, and using the fact that $w^{\Ls}(\ps)\approx 1/\sqrt{u}$ (see~(\ref{vlap})),
i.e.,
$w(s)\approx1/\sqrt{\pi s}$, we obtain the continuum limit of the response
\beq\label{eq:rnscal}
R_n(s,s+\tau)\approx\frac{1}{\pi\sqrt{2s\tau}}\e^{-\frac{(n-V\tau)^2}{2\tau}},
\eeq
which, again, is a solution of~(\ref{diffu}).
In contrast with the two-time correlation function, there is no anomalous aging regime for the response because the two times $s$ and $\tau$ are well decoupled.

In this regime, using~(\ref{dcsscal}) and~(\ref{eq:rnscal}), 
we find
\beqa\label{XT0}
X_n(s,s+\tau)^{-1}\approx
\frac{2 (s+\tau )}{2 s+\tau }\\+
\frac{\sqrt{\pi}\sqrt{ s \tau } 
(n-V(4 s +3 \tau) ) }{(2 s+\tau )^{3/2}}
\e^{\frac{s (n-V \tau )^2}{\tau (2 s+\tau )}} 
\mathop{\rm erf} \left[\frac{(n-V\tau)\sqrt{s}}{\sqrt{\tau(2s+\tau)}}\right].
\eeqa

\begin{figure}
\begin{center}
\includegraphics[angle=0,width=.9\linewidth]{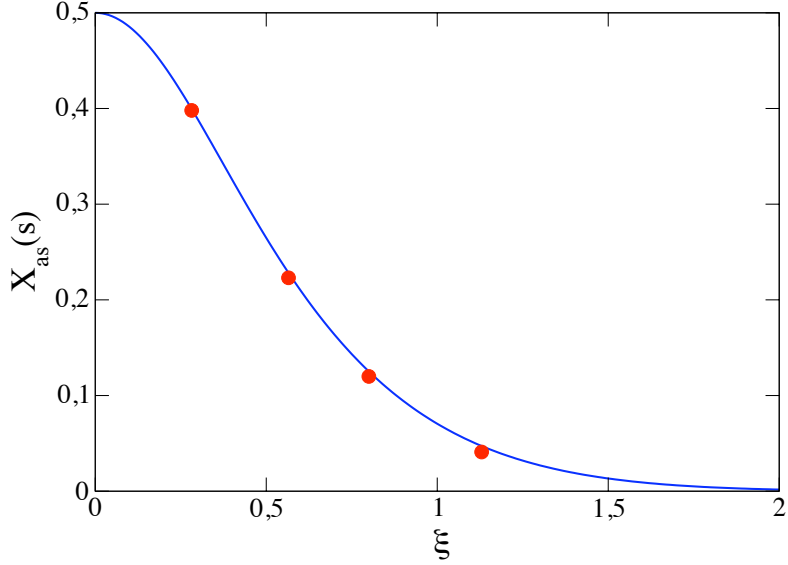}
\caption{\label{fig:Xas}
Asymptotic fluctuation-dissipation ratio $X_{\mathrm{ as}}(s)$ in the zero-temperature continuum limit against the scaling variable $\xi=V\sqrt{s}$.
Continuous line: analytical prediction~(\ref{Xscal}), dots: numerical integration of the equations for correlation and response. Successively, $s=32, V=0.05$ ($\xi=0.282$), $s=32,V=0.1$ ($\xi=0.565$), $s=16, V=0.2$ ($\xi=0.8$), $s=32, V=0.2$ ($\xi=1.13$).}
\end{center}
\end{figure}
Let us discuss some limiting cases of this expression.

\begin{enumerate}
\item
For $\tau=0$ we obtain
\beq
X_n(s,s)\approx 1.
\eeq
This result, which is known to hold in the symmetric case~\cite{gl2000}, proviso $s$ is large, which is indeed the case in the continuum limit considered here, is thus extended to the case of a non zero asymmetry.
\item
More generally, if $s\to\infty$ at fixed $\tau$, we have 
\beq
\lim_{s\to\infty}X_n(s,s+\tau)=\frac{1}{1-\sqrt{2\pi}V\sqrt{\tau}\e^{\frac{(n-V\tau)^2}{2\tau}}
\mathop{\rm erf}\left[\frac{n-V\tau}{\sqrt{2\tau}}\right]},
\eeq
hence, if $n=0$,
\beq
\lim_{s\to\infty}X_0(s,s+\tau)=\frac{1}{1+2g\big(V\sqrt{\tau/2}\big)},
\eeq
where $g(x)=\sqrt{\pi}\,\mathop{\e^{x^2}}x\mathop{\rm erf}x$.

\item
For $V$ and $n$ separately equal to zero the second term in~(\ref{XT0}) vanishes and we recover the result found for symmetric dynamics~\cite{gl2000,lip2000}
\beq
X_0(s,s+\tau)\approx\frac{2 s +\tau}{2(s+\tau)}.
\eeq
The same result holds for $n=V\tau$.
For these two cases, or more generally for $V=0$ and any value of $n$, letting $\tau\to\infty$ in~(\ref{XT0}), we obtain
\beq
X_{\mathrm{ as}}(s)=\frac{1}{2},
\eeq
i.e., $X_{\infty}=1/2$.

\item
For generic values of $n$ and $V$, the expression~(\ref{XT0}) can be positive or negative, and arbitrarily large.
When $\tau\to\infty$, we obtain the asymptotic scaling form
\beq\label{Xscal}
X_{\mathrm{ as}}(s)=\frac{1}{2+3g\big(V\sqrt{s}\big)}=
\frac{1}{2 +3 \sqrt{\pi}\,
\e^{V^2 s } V\sqrt{s}\,\mathop{\rm erf}\left[V\sqrt{s} \right]},
\eeq
which is a function of $V \sqrt{s}$ only, with no dependence on $n$.
For small values of the scaling variable $V\sqrt{s}$ we have
\beq
X_{\mathrm{ as}}(s)\approx \frac{1}{2}-\frac{3 V^2 s}{2}+\cdots,
\eeq
while it converges as $\e^{-V^2 s }$ to its limit value $X_\infty=0$.
The scaling form~(\ref{Xscal}) interpolates between $X_\infty=1/2$ for $V\sqrt{s}=0$ and $X_\infty=0$ for $V\sqrt{s}\to\infty$, as depicted in figure~\ref{fig:Xas}.

\end{enumerate}

As a conclusion, the non-trivial value of the zero-temperature limit fluctuation-dissipation ratio $X_\infty$ found for symmetric dynamics~\cite{gl2000,lip2000}, and characteristic of criticality, is lost in the presence of an asymmetry (except at the special point $n=V\tau$).
At zero temperature, under asymmetric dynamics, the system loses its critical character, yet keeping many of the characteristic features of a coarsening system.
More precisely, the two-time quantities, namely the correlation and response functions, decay exponentially for asymmetric dynamics, while they have power-law decay for symmetric dynamics.
On the other hand, the equal-time correlation function is insensitive to the presence of an asymmetry, and therefore all the statistics related to the growth of domains is unchanged.

\section{Discussion}

The model considered in the present work demonstrates how directedness, or the asymmetry of the dynamics, affect the properties of the original Glauber-Ising model, which appears as a special case of the model where dynamics is symmetric.

The two-spin equal-time correlation function turns out to be insensitive to the asymmetry, at any temperature.
Restricting to zero temperature, where the process can be mapped on a one-dimensional diffusion-annihilation reaction $A+A\to0$ with drift, $A$ representing a domain wall, this result is consistent with the analysis of former studies~\cite{gunter} which predict that density fluctuations of the $A$-particle do not depend on the bias, as long as the initial state is translationally invariant.

The effect of the asymmetry manifests itself through the emergence of a new scale, the asymmetry length scale $\xi_\mathrm{ asym}$.
In the stationary state, two regimes for the behaviour of the correlation, hence of the limit fluctuation-dissipation ratio, are observed.
These two regimes are separated in the velocity-temperature plane by a critical line $V=V_c=\sqrt{1-\gamma^2}$, or equivalently $\xi_{\eq}=\xi_\mathrm{ asym}$.
As long as this length scale is greater than the equilibrium correlation length $\xi_{\eq}$, or equivalently for values of the velocity less than the critical value $V_c$, the system departs weakly from the symmetric case, for which the fluctuation-dissipation theorem holds,
and the response and correlation functions have the same asymptotic decay laws.
The limit stationary fluctuation-dissipation ratio $X_{\infty,\stat}$, equal to the ratio of two amplitudes associated to correlation and response, is a function of temperature and velocity, taking values between 1, at $V=0$, and $0$, at $V=V_c$, and can be interpreted, in the spirit of~\cite{ck1}, as giving a measure of the effective temperature of the system.
Beyond the critical velocity $V_c$, i.e., when $\xi_{\eq}>\xi_\mathrm{ asym}$, the effect of the asymmetry becomes dominant, and the correlation function has a slower decay rate than that of the response function, hence the limit ratio $X_{\infty,\stat}$ vanishes.

At zero temperature, the system is coarsening, and therefore out of stationarity.
Yet a similar analysis predicts that, for $s$ fixed, and for any value of $V$, except $V=\pm1$, correlation and response have the same asymptotic decay laws at large temporal separation.
This can be explicitly seen in the continuum limit regime, where the fluctuation-dissipation ratio, in the limit $\tau\to\infty$, is a function of the scaling variable $V\sqrt{s}$ only, which vanishes in the limit $s\to\infty$.
At $V=\pm1$, corresponding to the totally asymmetric dynamics, the fluctuation-dissipation ratio asymptotically vanishes, when $\tau\to\infty$, even at finite $s$.

A possible extension of the present work would consist in performing the analysis of the behaviour of multispin two-time correlation and response functions, in order to investigate the role of the asymmetry, compared to what is known in the symmetric case~\cite{sollich1}.
At zero temperature, the two-time correlations of the energy have been studied in the past, in the framework of driven reaction-diffusion processes~\cite{stinch}.

\subsection*{Acknowledgments.}
It is a pleasure to thank J.-M. Luck for very fruitful discussions and for pointing ref.~\cite{luckmehta} to me.
I also thank O. Zaboronski for our conversations in the initial stage of this work and for sending me his notes on the free fermion methods applied to the directed Ising chain.
\newpage

\section*{Appendix}

In Laplace space the two-time correlation function $C_0(s,s+\tau)$ reads
\beqa\label{app:c0}
C_{0}^{\Ls\L}(u,v)
=\frac{\sqrt{(\ps+2)^2-4\gamma^2}}{\ps}
\frac{1}{2\gamma^2}I(a,b,V)
\eeqa
where
\beq
I(a,b,V)=\int_{0}^{2\pi}\frac{\d q}{2\pi}
\frac{1}{(a-\cos q)(b-\cos q+\i V\sin q)},
\eeq
with 
\beq
a=\frac{u+2}{2\gamma},\qquad b=\frac{v+1}{\gamma}.
\eeq

We are interested in the asymptotic behaviour of $C_0(s,s+\tau)$ at large $\tau$, hence
in the singularities of the integral $I(a,b,V)$ in the Laplace variable $v$, conjugate to $\tau$, or else in its singularities in the
variable $b$.
This integral is equal to
\beq\label{app:int}
\frac{b^2-a^2 y^2}
{\sqrt{a^2-1} \sqrt{b^2-y^2} \left[\sqrt{a^2-1}\left(b-a y^2\right)
- (a-b) \sqrt{b^2-y^2}\right]},
\eeq
where $y^2$ is a temporary compact notation for $1-V^2$.

For $V=0$ it has the simpler form
\beq
I(a,b,0)=\frac{a+b}{\sqrt{a^2-1}\sqrt{b^2-1}(\sqrt{a^2-1}+\sqrt{b^2-1})},
\eeq
which is singular at $b=b_1=1$.
For $V=1$, it has also a simpler form, reading
\beq\label{app:v1}
I(a,b,1)=\frac{1}{\sqrt{a^2-1}(b-a+\sqrt{a^2-1})},
\eeq
which is singular at $b=b_2=a-\sqrt{a^2-1}$.

In the general case, the singularities of the integral lie at $b=b_1=\sqrt{1-V^2}$ for $0\le V<1$, and 
at $b=b_2=a-V\sqrt{a^2-1}$ for $V_c\le V\le1$, where
$V_c=\sqrt{1-1/a^2}$, or equivalently $y=1/a$, is the value where $b_1=b_2$.
The first one corresponds to the vanishing of $\sqrt{b^2-y^2}$, the second one to the vanishing of the quantity inside the brackets in the denominator of~(\ref{app:int}), as we now show.

\begin{figure}
\begin{center}
\includegraphics[angle=0,width=.9\linewidth]{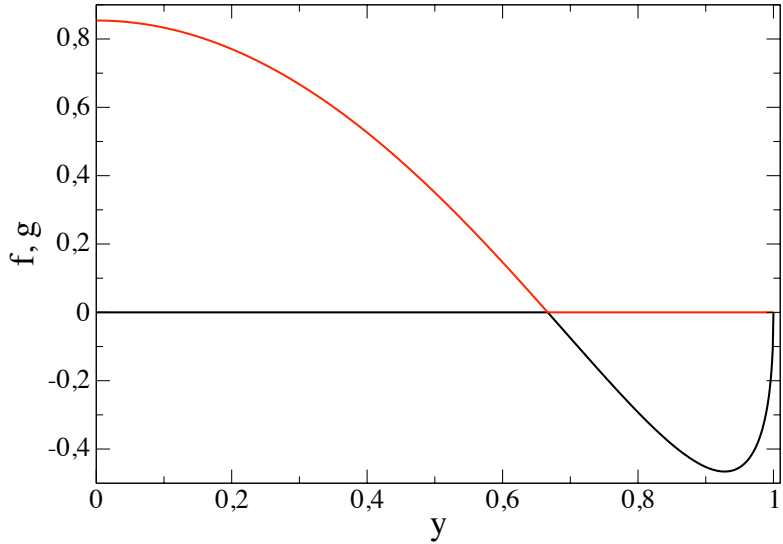}
\caption{\label{fig:app}
Representation of $f(b_{-};a,y)$ (in black) and $g(b_{-};a,y)$ (in red) as functions of $y$ for $a=1.5$.}
\end{center}
\end{figure}

We denote this quantity by
\beq
f(b;a,y)=\sqrt{a^2-1}\left(b-a y^2\right)
- (a-b) \sqrt{b^2-y^2},
\eeq
which is a function of $b$, depending on the two parameters $a$ and $y$.
In the present discussion we consider $a$ real ($a>1$).
Multiplying $f(b;a,y)$ by its conjugate, $g(b;a,y)=\sqrt{a^2-1}\left(b-a y^2\right)
+ (a-b) \sqrt{b^2-y^2}$, we find that the resulting expression vanishes for $b=\pm ay$ and for $b=b_{\pm}=a\pm V\sqrt{a^2-1}$. 
The singularities at $b=\pm ay$ are cancelled by the numerator of $I(a,b,V)$.
As for $b_{\pm}$, the following occurs.
For $a$ fixed, $f(b_{-};a,y)=0$ for $0\le y\le 1/a$, while this function is non zero for $1/a<y<1$.
Symmetrically, $g(b_{-};a,y)=0$ for $1/a\le y\le1$, while this function is non zero for $0<y<1/a$ (see Figure~\ref{fig:app}).
Finally, $f(b_{+};a,y)$ is positive for $0<y<1$, while $g(b_{+};a,y)=0$ on the same interval.

In summary, the only remaining singularity of $I(a,b,V)$ coming from $f(b;a,y)$ is at $b_{-}$, denoted by $b_2$ above, for $V_c\le V\le1$.
At $y=1/a$, or $V=V_c$, we have $b_1=b_2$.\footnote{The present discussion, done for the case $V>0$, is easily extended to the case $V<0$.}

Coming back to the two-time correlation function $C_0(s,s+\tau)$, we can conclude easily for the stationary case.
The stationary correlation function is the limit of~(\ref{app:c0}) multiplied by $u$, for $u\to 0$, which leads to the expression~(\ref{cfllstat}) in the text, corresponding to $a=1/\gamma$.
Hence the singularity at $b_2=a-V\sqrt{a^2-1}$ is explicit and no longer depends on the Laplace variable $u$.

For the case of zero temperature we proceed as follows.
Asymptotically, when $\tau\to\infty$, we have, at exponential order,
\beq\label{C0asymp}
C_0(s,s+\tau)\sim\e^{\tau g(u)},
\eeq
with $g(u)=b_2-1=u/2-V\sqrt{u^2/4+u}$.
We have to invert this expression with respect to the Laplace variable $u$.
Since $\tau$ is large, the saddle point method can be applied to the inverse Laplace transform, yielding $g'(u)=-s/\tau\to 0$.
The relevant solution $u_0=2(1/\sqrt{1-V^2}-1)$ corresponds to the minimum of $g(u)$.
Carried in~(\ref{C0asymp}) this yields, at exponential order,
\beq
C_0(s,s+\tau)\sim\e^{-\tau(1-\sqrt{1-V^2})},
\eeq
which corresponds precisely to the singularity $b_1$.
The conclusion is that for partially symmetric dynamics ($0<V<1$)
the asymptotic decay of the correlation function at large $\tau$ with $s$ fixed is governed by a single relaxation rate $\alpha_1=1-\sqrt{1-V^2}$.

Finally, at finite temperature, the very same analysis leads to the same conclusion
with $\alpha_1=1-\gamma\sqrt{1-V^2}$.

The case of totally asymmetric dynamics must be treated separately. 
We have from~(\ref{app:c0}) and (\ref{app:v1}), with $\gamma=1$,
\beq
C_{0}^{\Ls\L}(u,v)= 
\frac{1}{u(v-u/2+\sqrt{u^2/4+u})},
\eeq
yielding
\beq\label{app:C0L}
C_{0}^{\Ls}(u,\tau)=\frac{1}{u}\e^{\tau(\frac{u}{2}-\sqrt{\frac{u^2}{4}+u})}.
\eeq
In this case, the minimum of $g(u)=u/2-\sqrt{u^2/4+u}$ is reached at infinity as can be seen on the expression of $u_0$.
Expanding $g(u)$ for $u\to\infty$ gives $g(u)\approx -1+1/u$.
By inversion of $C_{0}^{\Ls}(u,\tau)\approx\e^{-\tau(1-1/u)}/u$ with respect to $u$ we finally  obtain
\beq
C_{0}(s,s+\tau)\approx\e^{-\tau}I_0(2\sqrt{s\tau})\sim\e^{-\tau+2\sqrt{s\tau}}.
\eeq

Let us mention another expression of the correlation for totally asymmetric dynamics,
\beq
C_0(s,s+\tau)=\sum_{m\ge0}C_m(s)G_m(\tau)=\e^{-\tau}\sum_{m\ge0}C_m(s)\frac{\tau^m}{m!}.
\eeq
Its Laplace transform with respect to $s$ yields back~(\ref{app:C0L}), using~(\ref{cnl}).


\newpage
\begin{thebibliography}{99}

\bibitem{glau} Glauber R G 1963 J. Math. Phys. {\bf 4} 297

\bibitem{langer} Langer J S in {\it Solids far from Equilibrium} 
Godr\`eche C ed. (Cambridge University Press, 1991)

\bibitem{bray} Bray A J 1994 Adv. Phys. {\bf 43} 357

\bibitem{cox} Cox J T and Griffeath D 1986 Ann. Probab. {\bf 14} 347

\bibitem{bray89} Bray A J 1989 J. Phys. A {\bf 22} L 67 

\bibitem{amar90} Amar J G and Family F 1990 Phys. Rev. A {\bf 41} 3258

\bibitem{bray97} Bray A J in {\it Nonequilibrium Statistical Mechanics
in One Dimension} Privman V ed. (Cambridge University Press, 1997)

\bibitem{prados97} Prados A Brey J J and S\'anchez-Rey B 1997 Europhys. Lett.
{\bf 40} 13

\bibitem{gl2000} Godr\`eche C and Luck J M 2000  J. Phys. A {\bf 33} 1151

\bibitem{lip2000} Lippiello E and Zannetti M 2000 Phys. Rev. E {\bf 61} 3369

\bibitem{malte} Henkel M and Pleimling M {\it Non equilibrium phase transitions Volume 2: Ageing and Dynamical Scaling Far from Equilibrium
} (Springer, 2010)

\bibitem{corb1} Corberi F Lippiello E and Zannetti M 2001 Eur. Phys. J. B {\bf 24} 359

\bibitem{sollich1} Mayer P and Sollich P 2004 J. Phys. A {\bf 37} 9

\bibitem{asep} For a review, see: Blythe R A and Evans M R 2007
J. Phys. A {\bf 40} R333

\bibitem{kls}
Katz S Lebowitz J L and Spohn H 1983 Phys. Rev. B {\bf 28} 1655
\nonum
Katz S Lebowitz J L and Spohn H 1984 J. Stat. Phys. {\bf 34} 497

\bibitem{zrp} For a review, see: Evans M R and Hanney T 2005 J. Phys. A {\bf 38} R195
\nonum
Godr\`eche C 2007 Lect. Notes Phys. {\bf 716} 261

\bibitem{lg2006} Luck J M and Godr\`eche C 2006 J. Stat. Mech. P08009 

\bibitem{gb2009} Godr\`eche C and Bray A J 2009 J. Stat. Mech. P12016

\bibitem{kls+} Zia R K P 2010 J. Stat. Phys. {\bf 138} 20

\bibitem{kun} K\"{u}nsch H R 1984 Z.Wahrscheinlichkeitstheorie
verw. Gebiete {\bf 66} 407 

\bibitem{stau} Lima F W S and Stauffer D 2006 Physica A {\bf 359} 423

\bibitem{ayyer} Ayyer A 2010 arXiv:1012.0875v1

\bibitem{gunter} Sch\"{u}tz G M 1995 J. Phys. A {\bf 28} 3405
\nonum
Sch\"{u}tz G M 1996 Phys. Rev. E {\bf 53} 1475
\nonum
Santos J E Sch\"{u}tz G M and Stinchcombe R B 1996 J. Chem. Phys. {\bf 105}  2399 

\bibitem{stinch} Grynberg M D and Stinchcombe R B 1995 Phys. Rev. Lett. {\bf 76} 851

\bibitem{baxter} Baxter R J {\it Exactly Solved Models in Statistical
Mechanics} (Academic Press, London, 1982)

\bibitem{luckmehta} Luck J M and Mehta A 2001 Europhys. Lett. {\bf 54} 573

\bibitem{ck1} Cugliandolo L F Kurchan J and Peliti L 1997 Phys. Rev. E {\bf 55} 3898

\bibitem{ck2} Cugliandolo L F and Kurchan J 1994 J. Phys. A {\bf 27} 5749

\bibitem{ck3} Cugliandolo L F Kurchan J and Parisi G 1994 J. Physique I {\bf 4} 1641

\bibitem{chate} Chatelain C 2003 J. Phys. A {\bf 36} 10739

\bibitem{fede} Ricci-Tersenghi F 2003 Phys. Rev. E {\bf 68} 065104(R) 

\bibitem{corb} Lippiello E Corberi F and Zannetti M 2005 Phys. Rev. E {\bf 71} 036104

\bibitem{diez} Diezemann G 2005 Phys. Rev. E {\bf 72} 011104

\bibitem{seif} Speck T and Seifert U 2006 Europhys. Lett. {\bf 74} 391

\bibitem{maes} Baiesi M Maes C and Wynants B 2009 Phys. Rev. Lett. {\bf 103} 010602

\bibitem{corb3} Corberi F Lippiello E Sarracino A and Zannetti M 2010 Phys. Rev. E {\bf81} 011124 

\bibitem{gl2000+} Godr\`eche C and Luck J M 2000 J. Phys. A {\bf 33} 9141

\bibitem{gl2002} Godr\`eche C and Luck J M 2002 J. Phys. C {\bf 14} 1589

\bibitem{calabrese} For a review, see: Calabrese P and Gambassi A 2005 J. Phys. A {\bf 38} R133

\bibitem{sollich2} Garriga A Sollich P Pagonabarraga I and Ritort F 2005 Phys. Rev. E {\bf 72} 056114

\bibitem{sollich3} Annibale A and Sollich P 2006 J. Phys. A {\bf 39} 2853

\bibitem{corb2} Corberi F Lippiello E and Zannetti M 2004 J. Stat. Mech. P12007 

\bibitem{x} Barrat A 1998 Phys. Rev. E {\bf 57} 3629 
\nonum
Berthier L Barrat J L and Kurchan J 1999 Eur. Phys. J B {\bf 11} 635


\end{thebibliography}
\end{document}